\documentclass{amsart}
\usepackage[english]{babel}
\usepackage{geometry}
\usepackage{textcomp,xcolor}
\usepackage{subfigure}
\usepackage{epsfig}
\usepackage{multirow}
\usepackage{bm}
\usepackage{graphics}
\usepackage{amsmath}

\newtheorem{definition}{Definition}
\newcommand*\dif{\mathop{}\!\mathrm{d}}
\newcommand{\vect}[1]{\boldsymbol{\mathbf{#1}}}

\makeatletter
\def\ps@pprintTitle{%
\let\@oddhead\@empty
\let\@evenhead\@empty
\def\@oddfoot{}%
\let\@evenfoot\@oddfoot}
\makeatother

\begin{document}

%
\title[Dimension reduction in heterogeneous neural networks]{Dimension reduction in heterogeneous neural networks: generalized Polynomial
Chaos (gPC) and ANalysis-Of-VAriance (ANOVA)}


\author{Minseok Choi}
\address[Minseok Choi and Tom Bertalan]{Department of Chemical and Biological Engineering,
	Princeton University, Princeton, NJ 08544, USA}

\author{Tom Bertalan}
\author{Carlo R. Laing}
\address[Carlo R. Laing]{Institute of Natural and Mathematical Sciences,
	Massey University, Auckland, New Zealand}

\author{Ioannis G Kevrekidis}
\address[Ioannis G Kevrekidis]{Department of Chemical and Biological Engineering
and Program in Applied and Computational Mathematics,
	Princeton University, Princeton, NJ 08544, USA}




\begin{abstract}
We propose, and illustrate via a neural network example, two different approaches to coarse-graining large heterogeneous
networks.
Both approaches are inspired from, and use tools developed in, methods for uncertainty quantification in systems
with multiple uncertain parameters - in our case, the parameters are {\em heterogeneously} distributed
on the network nodes.
The approach shows promise in accelerating large scale network simulations as well as coarse-grained 
fixed point, periodic solution and stability analysis.
We also demonstrate that the approach can successfully deal with \emph{structural} as well as
\emph{intrinsic} heterogeneities.
\end{abstract}

\maketitle


%
%
\section{Introduction}
\label{sec:intro}
Systems of coupled {\em identical} oscillators can often be studied exploiting this special symmetry
(invariance to permuting their identities~\cite{ashwin92});
yet most realistic systems possess some form/degree of heterogeneity, and thus studying the influence
of this heterogeneity on dynamics is of crucial importance.
While for a small number of oscillators
the dynamics of each and every one can be easily simulated, for larger networks this becomes impractical, particularly
if one is interested in {\em typical} behaviour of similar networks, not just the behavior of a single, particular
network realization.
Thus techniques for dimension reduction, i.e.~faithfully representation of a heterogeneous network
by a lower-dimensional dynamical system, are useful in the dynamic/parametric study of such networks.

In this paper we demonstrate the use of two such dimensionality reduction techniques for
different heterogeneous networks of coupled model neurons.
The first network we consider is all-to-all coupled, but four of the physiological parameters associated
with the dynamical mechanisms occurring within each neuron are heterogeneous. 
The term that embodies all-to-all coupling of the neurons is then approximated
by a four-dimensional integral over these heterogeneous parameters.
We approximate this integral using the ANalysis-Of-VAriance (ANOVA) method and expand
the instantaneous states of the neurons in polynomials in the four heterogeneous parameters.
A small number of time-dependent coefficients for these polynomials constitute
the variables of  a reduced model for the network.
We demonstrate the computational efficiency of this reduction with several computations within
the equation-free framework (e.g. \cite{kevrekidis03,theodoropoulos00}).

The second network we consider is both \emph{intrinsically heterogeneous} (a
physiological parameter associated with the individual neuron dynamics is different for each neuron)
as well as \emph{structurally heterogeneous} (because the neurons are connected in a
nontrivial way). 
We observe that the state of each neuron can be
accurately expressed as a sum of polynomials in both the intrinsic
heterogeneity parameter and a neuron's degree (number of connections) in the network.
This is a generalized Polynomial Chaos (gPC) approach, and the polynomials are orthogonal
with respect to a density that depends on the probability distribution
of the heterogeneous intrinsic parameter 
as well as the degree distribution of the network. 
A small
number of the coefficients of these polynomials again helps construct an accurate 
reduced model of the network dynamics.

The model is presented in Sec.~\ref{sec:model} and we then briefly review both ANOVA
and the gPC method. Numerical examples are given in Sec.~\ref{sec:num} and we conclude with a discussion in
Sec.~\ref{sec:conclusion}.

%
%
\section{The model}
\label{sec:model}

We consider a network of model neurons previously studied as a model for rhythmic
oscillations in the pre-B{\"o}tzinger complex~\cite{carlo12,rubin02}:
\begin{subequations} \label{eqn:botzinger}
	\begin{align}
  C \frac{dV_i}{dt} & =  - g_{Na} m (V_i) h_i (V_i-V_{Na})
  	- g_l (V_i - V_l) + I_{syn}^i + I_{app}^i, \label{subeqn:botzinger1} \\
  \frac{dh_i}{dt} & =  \frac{h_{\infty} (V_i) - h_i}{\tau (V_i)}
	  \label{subeqn:botzinger2}
\end{align}
\end{subequations}
for $i=1,...,N$, where
\begin{equation} \label{eqn:botzinger3}
	I_{syn}^i = \frac{g_{syn}(V_{syn} - V_i)}{N} \sum_{j=1}^N A_{i,j} s(V_j).
\end{equation}
Here $V_i$ is the membrane potential of neuron $i$, and $h_i$ is a channel
state variable for neuron $i$ that governs the inactivation of
persistent sodium. 
The first and second term of the right hand side in Equation
\eqref{subeqn:botzinger1} is a persistent sodium current and passive
leakage current, respectively and $g_{Na}, V_{Na}, g_l, V_l$ are
corresponding nominal parameters \cite{butera99models_a,butera99models_b}.
Equation~\eqref{eqn:botzinger} was derived from the models in
Butera \textit{et al.} \cite{butera99models_a,butera99models_b}  by blocking currents
responsible for action potentials; Rubin \cite{rubin06bursting} considered a similar
model with $N=2$, and Dunmyre and Rubin \cite{rubin10optimal} considered
synchronization in the case $N=3$.
The various functions involved in the model equations are
as follows:
\begin{eqnarray}
	s(V) &= &\frac{1}{1 + \exp(-(V+40)/5)},	\label{eqn:fcn_s} 	\\
	\tau(V) &= & \frac{1}{\epsilon \cosh((V+44)/12)},
		\label{eqn:fcn_t} \\
	h_{\infty}(V) &= &\frac{1}{1 + \exp((V+44)/6)},	\label{eqn:fcn_h} 	\\
	m(V) &= &\frac{1}{1 + \exp(-(V+37)/6)}.	\label{eqn:fcn_m} 	
\end{eqnarray}
The functions $\tau(V), h_\infty(V)$ and $m(V)$ are a standard part of
the Hodgkin-Huxley formalism~\cite{hassard78}, and synaptic communication is
assumed to act instantaneously through the function $s(V)$.
The neurons are coupled through a synaptic current $I_{syn}^i$ for
$g_{syn} \neq 0$ where $A_{ij}$ is a symmetric adjacency matrix, i.e.
$A_{ij}=1$ if neuron $i$ and $j$ are connected, and $A_{ij}=0$
otherwise. 
A previous study considered only all-to-all coupled networks~\cite{carlo12}, but we will
consider a more structured network in Sec.~\ref{sec:struc} below.
We denote the degree of $i$-th neuron (its number of
neighbors) by $\kappa_i$, i.e. $\kappa_i=\sum_{j \neq i}^N A_{ij}$.

It was shown in \cite{carlo12,rubin02} that if the values of the
applied currents $I_{app}^i$ are uniformly distributed in a certain interval, 
synchronous behavior is observed after a transient, \textit{i.e.} all neurons
oscillate periodically with the same period, although the
heterogeneity in the $I_{app}^i$ means that each neuron follows a
slightly different periodic orbit in its own $(V,h)$ phase space. 
It appears that (asymptotically in time) the values of the $V_i$ and $h_i$ vary smoothly as a function of
the heterogeneous parameter $I_{app}^i$. 
This observation lead to the continuum
limit of Equations (\ref{eqn:botzinger}):
\begin{subequations} \label{eqn:botzinger_cont}
\begin{align}
  C \frac{\partial V(\mu,t)}{\partial t} & = - g_{N_a} m (V
    ({\mu}, t)) h ({\mu}, t)  (V ({\mu}, t) - V_{N_a})
  - g_l (V ({\mu}, t) - V_l) + I_{syn} + I_{app}
    \label{subeqn:botzinger4} \\
  \frac{\partial h(\mu,t)}{\partial t} & = \frac{h_{\infty} (V ({\mu},
      t)) - h ({\mu}, t)}{\tau (V ({\mu}, t))} \label{subeqn:botzinger5}
\end{align}
\end{subequations}
where $I_{app}$ is parameterized as $I_{app} = I_m + I_s \mu$ with
$\mu$ being a uniform distribution on $[-1,1]$, \textit{i.e.}
$I_{app}$ follows a uniform distribution on $[I_m-I_s, I_m+I_s]$ and
\begin{equation} \label{eqn:Isyn}
	I_{syn}(\mu,t) = g_{syn}(V_{syn} - V(\mu,t)) \int_{-1}^1
	s(V(\mu,t)) p(\mu) d\mu.
\end{equation}

Note that $p(\mu)$ is a probability density function for $\mu$,
\textit{i.e.} $p(\mu) = 1/2$ for $-1\le \mu \le 1$. 
In this limit
$V_i(t)$ and $h_i(t)$ become the functions $V(\mu,t)$ and $h(\mu,t)$,
respectively.
The results for $N \rightarrow \infty$ should provide a good approximation to the
behavior seen when $N$ is large but finite, as we expect it to be. 
Rubin and Terman
\cite{rubin02} first introduced the continuum limit, their contribution
being largely analytical. 
Laing et al.~\cite{carlo12}
presented a computationally
efficient way to describe the heterogeneous network by applying techniques
widely used in the uncertainty quantification (UQ) community known as
generalized Polynomial Chaos and the associated stochastic collocation method
(SCM) \cite{xiu05pcm,xiu02gpc,ghanem03}.
These methods are high-order accurate, in fact exponentially accurate, but
suffer when the dimensionality of the parametric space increases; this is known
as the so-called \textit{curse of dimensionality}. 
Sparse grids
techniques have greatly alleviated this problem by utilizing the
smoothness of the function in low to moderate ``heterogeneity dimensions"
\cite{foo08mepcm,griebel98sparse}.
However, the complexity estimate of sparse grids still
depends heavily on the dimension and on the regularity of the
functions being integrated. 
To push the dimensionality barrier higher, several
methods have been introduced in the UQ commnunity; one of them is the ANOVA method, which
will be described in Sec.~\ref{subsec:multi_hetero}
for a case in which
there are multiple heterogeneous physiological (intrinsic to each neuron) parameters.
In the subsections~\ref{subsec:pc} and~\ref{subsec:anova} we briefly review the gPC and ANOVA methods;
see \cite{foo08mepcm,xiu02gpc,choi12anova,ghanem03} for more details.

\subsection{Polynomial Chaos as a low-dimensional representation}
\label{subsec:pc}
The Polynomial Chaos (PC) method is widely
used in the UQ community \cite{xiu02gpc,ghanem03}.
The method has also been applied successfully to coarse-graining
the dynamics of heterogeneous networks, for which some parameters \emph{intrinsic} to each neuron are
distributed in a prescribed way~\cite{carlo12,moon05}.
The PC expansion involves representing the state variable $\bm
X=(x_1,...,x_n)$ as a weighted
series of orthogonal basis functions (polynomials) of the heterogeneous
parameters $\bm \xi=(\xi_1,...\xi_m)$:
\begin{equation} \label{eqn:gpc_representation}
	\bm X(t; \bm \xi) = \sum_i \bm \alpha_i(t) \bm
				\Psi_i(\bm \xi)
\end{equation}
where $\bm \Psi_i(\bm \xi)$ is the $i$-th basis function and the
$\bm \alpha_i(t)$ are PC coefficients.
Conversely, the coefficients $\bm
\alpha_i$ can be recovered by the projection on the basis $\Psi_i(\bm
\xi)$ due to the orthonormality of the basis functions
\begin{equation} \label{eqn:gpc_coeff}
    \bm \alpha_i = \langle \bm X,\, \bm \Psi_i \rangle
    \equiv \int X(\bm \xi) \Psi_i(\bm \xi) dP(\bm \xi)
\end{equation}
where the inner product $\langle\cdot,\,\cdot\rangle$ is defined by integration with respect to
the underlying measure $dP(\bm \xi)$.

Assuming independence of the distributions of the heterogeneous parameters, $\bm \Psi_i$ can be
separated into a tensor product of independent scalar polynomial bases
$\bm \Psi_i(\bm \xi) = \prod_{k=1}^m \Psi_{i_k} (\xi_k)$.
For well-known distributions such as uniform or normal, there are
corresponding PC basis functions: Legendre or
Hermite polynomial, respectively. 
In the case of arbitrarily
distributed heterogeneous parameters, a PC basis can be constructed numerically \cite{wan06}.
With a basis chosen, the system of coupled ODEs for $\bm X$
\begin{equation} \label{eqn:X}
	\frac{d\bm X(t;\omega)}{dt} = \bm f(\bm X; \bm \xi)
\end{equation}
(of which~\eqref{eqn:botzinger} is a specific example)
can be recast as a system of ODEs for the PC coefficients $\bm
\alpha_i$ via the Galerkin method
\begin{equation}
	\frac{d \bm \alpha_j}{dt} = \left < \bm f \left(\sum_i \bm \alpha_i \bm
	\Psi_i \right), \bm \Psi_j \right>,
\label{eq:dalpha}
\end{equation}
where the orthogonality of the basis functions is exploited.

A computational task involving simulating each neuron in a system such as Equation
(\ref{eqn:botzinger}) is too complicated if the number of neurons is large, and
hence an accurate coarse-grained description is useful (if it exists). 
It turns out that PC coefficients $\bm \alpha_i$ serve well as coarse-grained
descriptors of a system like~\eqref{eqn:X} with heterogeneous
parameters~\cite{carlo12,moon05}. 
Note that the number of coefficients is usually much less
than the number of variables in Equation (\ref{eqn:X}). 
This model reduction, as we will show below, 
allows us to perform a number of coarse-grained modeling tasks
such as accelerated simulation via Coarse Projective Integration
(CPI) or accelerated limit cycle computation, accompanied by coarse-grained
stability analysis~\cite{bold07,laing08,carlo12,moon15,moon05}.

Coarse-graining this model requires the computation of two high-dimensional
integrals--the coupling integral \eqref{eqn:Isyn},
and the inner product \eqref{eqn:gpc_coeff}.
To this end, we introduce the ANOVA method in the following subsection.

\subsection{ANOVA}
\label{subsec:anova}
ANOVA is widely used as a statistical method
to test differences between two or more means~\cite{hoeffding,fisher}.
The same idea can be used for the
interpolation and integration of high dimensional problems as well as analyzing stochastic simulations.
\cite{foo10mepcm,sobol01}. 
Consider an integrable function
$f(\bm x)$, $\bm x=(x_1,x_2,\cdots,x_N)$ defined in $I^N=[0,1]^N$.
The ANOVA representation for $f(\bm x)$ is as follows:
\begin{definition}
	The representation of $f(\bm x)$ in a form
\begin{equation}\label{anova1}
f(\bm x)=f_0 + \sum_{s=1}^N\sum_{j_1<\cdots<j_s} f_{j_1\cdots j_s}(x_{j_1},\cdots,x_{j_s})
\end{equation}
or equivalently
\begin{equation}\label{anova0}
f(\bm x)=f_0 + \sum_{1\leq j_1\leq N} f_{j_1}(x_{j_1}) + \sum_{1\leq j_1<j_2\leq N} f_{j_1,j_2}(x_{j_1}, x_{j_2}) +
\cdots + f_{1, 2, \cdots, N}(x_{1}, x_{2}, \cdots, x_{_N})
\end{equation}
is called the ANOVA decomposition of $f(\bm x)$, if
\begin{equation}
\label{anova2}
	f_0=\int_{I^N} f(\bm x) \dif\mu(\bm x),
\end{equation}
and
\begin{equation}
\label{anova3}
	\int_{I} f_{j_1\cdots j_s}\dif\mu(x_{j_k})=0 \quad\mbox{for}\quad 1\leq k \leq s.
\end{equation}
We call $f_{j_1}(x_{j_1})$ the first-order
term (or first-order component function), $f_{j_1,j_2}(x_{j_1,j_2})$ the second-order term (or
second-order component function), etc.
\end{definition}
The terms in the ANOVA decomposition are computed as follows:
\begin{equation}\label{eqn:anovaterm}
f_{_S}=\int_{I^{N-|S|}} f(\bm x)\dif\mu(\bm x_{_{S^c}})-\sum_{T\subset S}f_{_T}(\bm x_{_T}),
\end{equation}
where $S=\{j_1,j_2,\cdots,j_s\}$, $|S|$
is the number of elements in $S$, $T=\{i_1,i_2,\cdots,i_t\}$ is a
subset of $S$, \textit{i.e.} $\{i_1,i_2,\cdots,i_t\} \subset
\{j_1,j_2,\cdots,j_s\}$ and $f_T=f_{i_1, i_2, \cdots, i_t}(x_{i_1},
x_{i_2}, \cdots, x_{i_t})$.

An important property of the ANOVA decomposition of $f$ is that the variance of
$f$ is the sum of the variances of all the ANOVA terms except $f_0$:
\begin{equation} \label{anovavar}
\sigma^2(f)=\sum_{s=1}^N\sum_{|S|=s}\sigma^2(f_{_S}),\quad\sigma^2(f_S) = \int_{I^N} f_S^2 d \mu (x),
\end{equation}
or equivalently:
\[\sigma^2(f)=\sum_{1\leq j_1\leq N}\sigma^2(f_{j_1})+\sum_{1\leq j_1<j_2\leq N}\sigma^2(f_{j_1, j_2})
 + \cdots + \sigma^2(f_{1,2,\cdots, N}).\]
Computing the ANOVA decomposition, i.e. the constant term
and high-order terms from Equations (\ref{anova2}) and
(\ref{eqn:anovaterm}) respectively, can be very expensive for high
dimensional problems or complicated functions $f(\bm x)$. 
One therefore uses the Dirac measure
instead of the Lebesgue measure in integrations, i.e.,
$\dif\mu(\bm x)=\delta(\bm x-\bm c)\dif\bm x,\, \bm c\in I^N$.
The point ``$\bm c$" is called the ``\textbf{\textit{anchor point}}" and this method is called
``\textbf{\textit{anchored-ANOVA}}". 
Then the (approximate) evaluation of the
integral that appears in the first term of the right hand side of
Equation (\ref{eqn:anovaterm}) becomes much easier. 
For example,
for the constant term and first-order term we have
\begin{eqnarray}
	f_0 &= & f({\bm c}) \\
	f_j(x_j) &= & f(c_1,...,c_{j-1},x_j,c_{j+1},...,c_N) - f_0,	\quad
    j=1,...,N \label{eqn:anova_firstorder}.
\end{eqnarray}
Note also that Equation (\ref{eqn:anovaterm}) implies that the
$|S|$-order terms can be constructed recursively from all ANOVA terms
whose orders are less than $|S|$.

For numerical purposes we approximate $f(\bm x)$ by all ANOVA terms
whose degrees are less than or equal to $\nu$:
\begin{equation}\label{eqn:anova4}
f(\bm x)\approx f_0 + \sum_{j_1\leq N} f_{j_1}(x_{j_1}) + \sum_{j_1<j_2\leq N} f_{j_1,j_2}(x_{j_1},
x_{j_2}) + \cdots + \sum_{j_1<j_2<\cdots<j_{\nu}\leq N} f_{j_1,j_2,\cdots,j_{\nu}}(x_{j_1}, x_{j_2},
\cdots x_{j_{\nu}}).
\end{equation}
Here $N$ is called \textit{nominal dimension}, and $\nu$ is called
the \emph{truncation} or \emph{effective dimension}. 
If $\nu$ is low, then this type
of approach, i.e. approximating the $N$-dimensional problem into a
series of lower-dimensional problem, can greatly alleviate the
computational burden. 
For example, let us consider the integration of
the function $\int f(\bm x) d\bm x$
, e.g., $f$ here can be the integrand in Equation (\ref{eqn:Isyn}) or
Equation (\ref{eqn:gpc_coeff}).
Since the integration is a linear
operator, the integral can be approximated by the sum of
integrals of ANOVA terms, i.e.
\begin{equation} \label{eqn:anova_integral}
	\int_{I^N} f(\bm x) d\bm x \approx \int_{I^N} f_0 d\bm x +
	\sum_{s=1}^\nu \sum_{j_1<\cdots<j_s} \int_{I^N} f_{j_1\cdots
	j_s}(x_{j_1},...,x_{j_s}) d\bm x.
\end{equation}
Then, the N-dimensional integration problem becomes much lower
dimensional (up to $\nu$ assuming $\nu \ll N$) integration, where we
can use collocation methods such as those involving Gaussian quadrature and weights.
Consider the first-order term $f_1(x_1)$ for instance. 
Let $\bm c_{-1}
= (c_2,c_3,...,c_N)$ and $(q_1^j, w^j)_{j=1}^\mu$ be the
quadrature points and corresponding weights for integration along the
first dimension, with $\mu$ being the number of quadrature points. 
Then, the integration of $f_1(x_1)$ can be approximated by
\begin{equation}
	\int_{I^N} f_1(x_1) d\bm x \approx \sum_{j=1}^\mu f_1(q_1^j) w^j
	= \sum_{j=1}^\mu (f(q_1^j,c_2,c_3,...,c_N) - f_0) w^j.
\end{equation}
See~\cite{foo10mepcm,choi12anova} for more details.
In \cite{choi12anova}, the authors applied the ANOVA
method for a stochastic incompressible flow problem with a nominal
dimension of parametric space up to 100 but with an effective
dimension of 2 as an efficient dimension-reduction technique.  
In Sec.~\ref{subsec:multi_hetero} below we
will demonstrate the use of ANOVA to approximately describe coupled neuronal networks
that have multiple independent heterogeneous parameters.

%
%
\section{Numerical examples}
\label{sec:num}
In this section, two cases are presented to illustrate the gPC and ANOVA
methods to model the effect of multiple heterogeneous parameters.
In the first case we model four distinct heterogeneous
parameters in order to demonstrate the ANOVA method:
$I_{app}, g_{N_a}, V_{syn}$ and $V_{N_a}$ are all assumed to be uniformly distributed.
For simplicity we do not assume \textit{structural} heterogeneity, i.e.
neurons are all-to-all coupled yielding $A_{ij}=1$
for all $i,j$; the case of simultaneous intrinsic \emph{and} strctural heterogeneity
will be discussed next.
After comparing the ANOVA method with sparse grids or the ``direct" Monte Carlo
(MC) method, we perform a number of coarse-grained modeling tasks such
as Coarse Projective Integration and coarse-grained stability analysis.

We then consider the network of neurons to be heterogeneous in the following sense:
neuron $i$ has an applied current $I_{app}^i$, which is referred
to as an $\textit{intrinsic}$ heterogeneity, and a
degree $\kappa_i$, which is referred to as a \textit{structural} heterogeneity.
(The $I_{app}^i$ are not all equal, and neither are the $\kappa_i$.)
The results suggest that the techniques used here may be also applicable to
this type of network.

\subsection{Case I: Multiple heterogeneous parameters}
\label{subsec:multi_hetero}
We consider the case where there exist four heterogeneous parameters:
$I_{app}, g_{N_a}, V_{syn}$ and $V_{N_a}$ are all independently and uniformly distributed.
Each of these four parameters can be parameterized by their mean and half-width,
together with the standard uniform
distribution $\xi_i, i=1,2,3,4$, which we denote by $\xi_i \sim U[-1,1]$, and whose
probability distribution function is $p(\xi_i) = \frac{1}{2}$ for $-1
\le \xi_i \le 1$. 
For example, if $I_{app} \sim U[17.5,32.5]$, then it
is parameterized as $I_{app} = E[I_{app}] + h(I_{app}) \xi$ where
$E[I_{app}]=25$ and $h(I_{app})=7.5$ are the mean and half-width of $I_{app}$,
respectively, and $\xi$ is the standard uniform distribution.
Then, as
mentioned in the above section, the continuous variables
$V$ and $h$ become a function of these $\xi_i$'s as well as time
$t$ as $V(t;\xi_1,\xi_2,\xi_3,\xi_4)$ and
$h(t;\xi_1,\xi_2,\xi_3,\xi_4)$, respectively and the sum in Equation
(\ref{eqn:botzinger3}) is represented by the integral
\begin{equation} \label{eqn:int_s}
    \int_{\Omega^4} s(V(t;\bm \xi)) p(\bm \xi) \dif\xi_i
\end{equation}
where $\bm \xi=(\xi_1,\xi_2,\xi_3,\xi_4), p(\bm \xi) = \prod_{i=1}^4
p_i(\xi_i)$  and $\Omega = [-1,1]$.
In stochastic collocation or sparse grid methods this integral is
approximated as the sum of the function evaluated at the collocation
points multiplied by their corresponding weights; see
\cite{carlo12,xiu05pcm} for more detail. 
In ANOVA methods, we first
approximate the function $s(V(t;\vec{\xi}))$ by its ANOVA terms whose
orders are less than $\nu$ as in Equation (\ref{eqn:anova4}). 
Then
the integral of a high-dimensional function is represented by the integral of a series of
low-order functions, which can be easily computed by
standard numerical integration techniques. 
For example, assume that
$\nu=1$. 
Then the ANOVA approximation of $s$ denoted by $s_A$ is as follows:
\begin{equation}
    s(V(t;\bm \xi)) \approx
        s_{A}(V(t;\bm \xi)) = s_0 + \sum_{j=1}^4 s_j(\xi_j)
\end{equation}
where $s_j(\xi_j)$ is given in Equation (\ref{eqn:anova_firstorder}). 
For example,
for $j=2$, $s_2(\xi_2) = s(V(t;c_1,\xi_2,c_3,c_4))$ for an anchor
point $\bm c = (c_1,c_2,c_3,c_4)$. 
Then the integral in Equation
(\ref{eqn:int_s}) is computed as the sum of the integral of the constant
term and the first-order ANOVA terms, which are readily computable:
\begin{equation}
    \int_{\Omega^4} s(V(t;\bm \xi)) p(\bm \xi) \dif\xi_i
\approx \int_{\Omega^4} s_A(V(t;\bm \xi)) p(\bm \xi) \dif\xi_i = E[s_0] + \sum_{j=1}^4 E[s_j(\xi_j)]
\end{equation}
where $E[f]$ is the expectation operator of $f$ with respect to the
probability measure $p(\bm \xi)$.

\begin{figure}[htbp]
	\centering
	\subfigure{
		\includegraphics[width=0.45\textwidth]{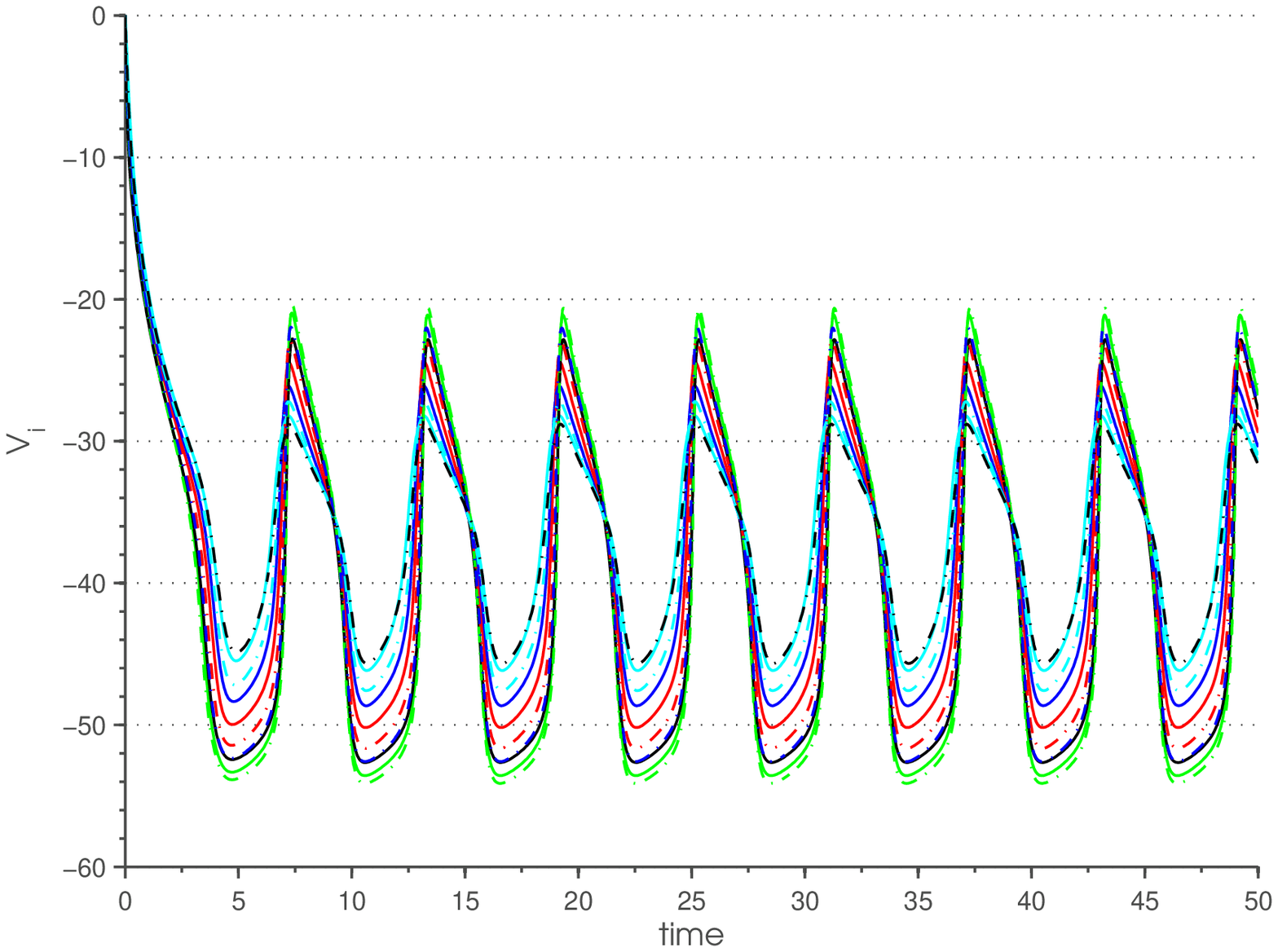} 
		\includegraphics[width=0.45\textwidth]{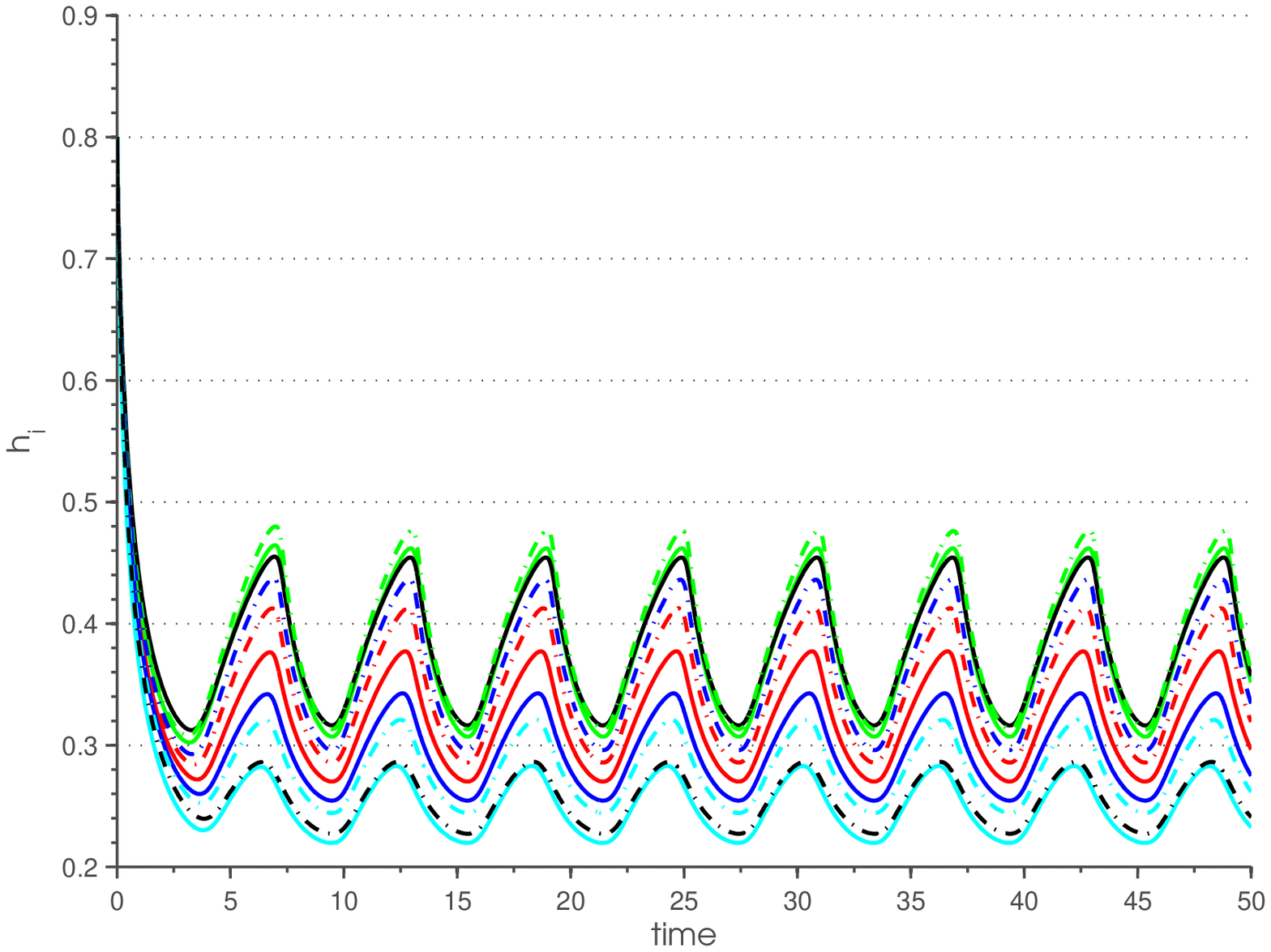}
	} 
	\subfigure{
		\includegraphics[width=0.45\textwidth]{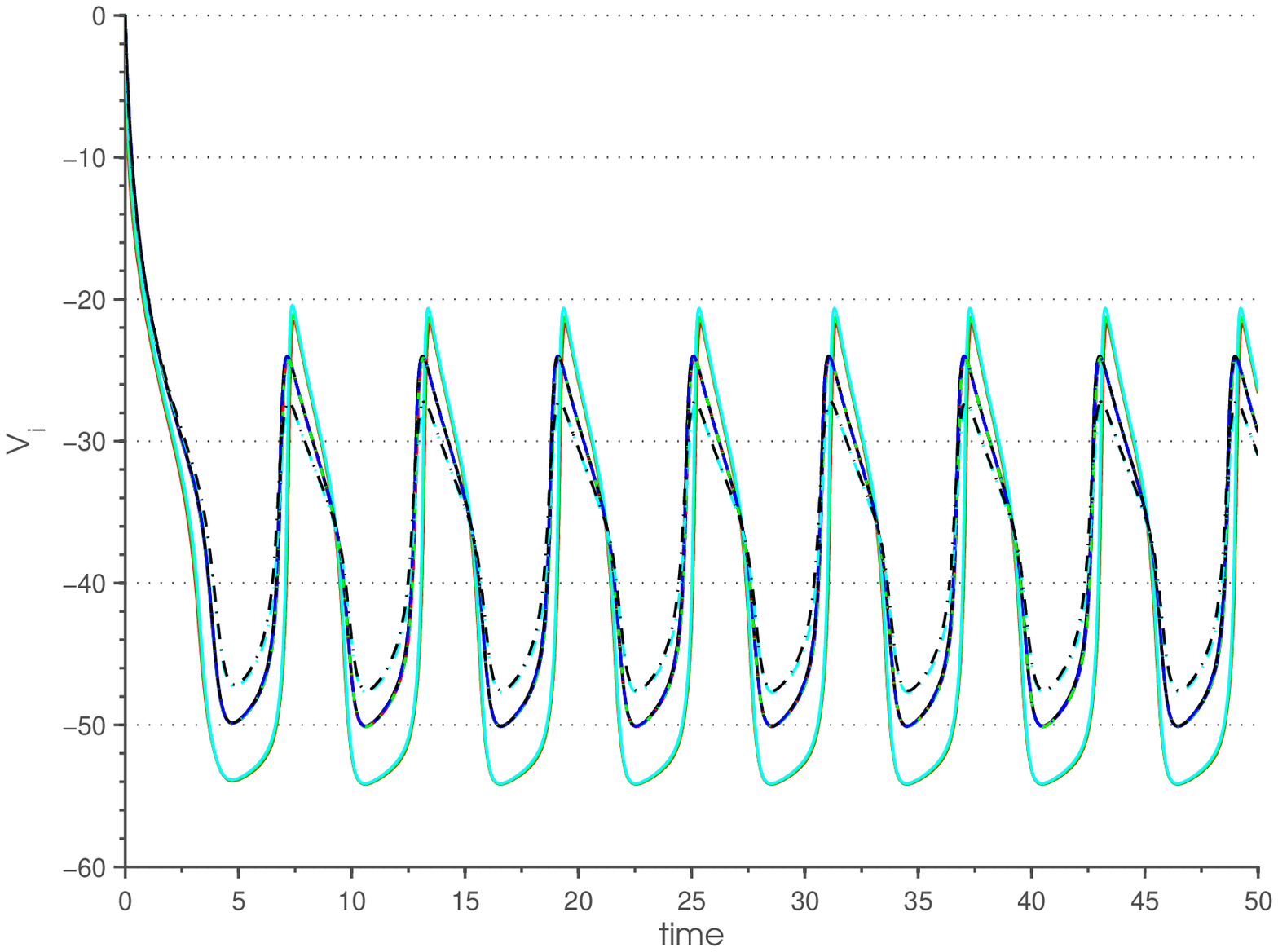} 
		\includegraphics[width=0.45\textwidth]{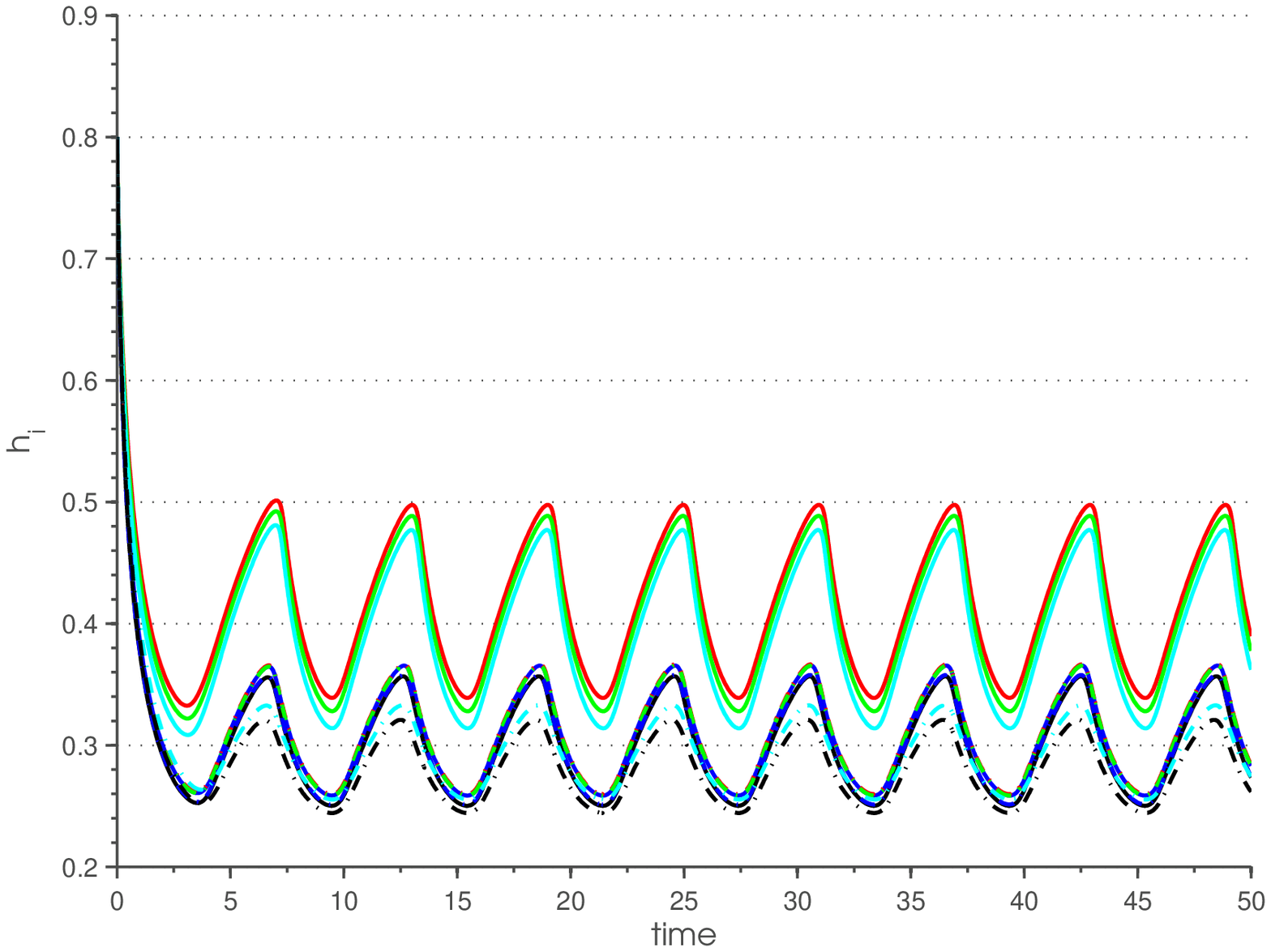}
	} 
	\caption{Solutions of Equations (\ref{subeqn:botzinger1}) and
    (\ref{subeqn:botzinger2}) when there are four heterogeneous parameters
	and samples come from sparse grids (Top) and ANOVA (bottom) with parameters given in Table
    \ref{tbl:sparse_anova}. \textit{Left:} $V_i$ as functions
	of time. \textit{Right:} $h_i$ as functions of time. Different line
	colors correspond to different neurons (only ten neurons are shown) and they show that neurons with different
	parameters behave differently.}
	\label{fig:samples_V_h}
\end{figure}

All four heterogeneous parameters here follow a uniform distribution:
$I_{app}$ on $[17.5,32.5]$, $V_{syn}$ on [-1,1], $V_{Na}$ on [49,51],
and $g_{Na}$ on $[2.55,3.05]$.
The other parameters are given as follows:
\begin{equation}
  V_{N_a} = 50, \quad V_{syn} = 0, \quad g_{syn} = 0.3,
  \quad g_l = 2.4, \quad V_l = - 65, \quad \varepsilon = 0.1,
  \quad C = 0.21. \nonumber
\end{equation}
The parameters for sparse grids and ANOVA are shown in Table~\ref{tbl:sparse_anova}.
We also consider a direct Monte Carlo (MC) method with 10,000 points (i.e. 10,000
all to all coupled neurons) as a
reference solution.
Note that both sparse grids and ANOVA methods MC are non-intrusive
methods, hence given the sampling
(or collocation) points, we solve deterministic problems. 
Figure \ref{fig:samples_V_h} shows the behavior of the $V_i$ and $h_i$ corresponding
to $10$ samples from the sparse grids in Table \ref{tbl:sparse_anova}.

\begin{table}[h]
	\centering
  \caption{In sparse grids, the number of collocation points is determined by
  the level, i.e. the higher the level the more points. In the ANOVA method,
  ${\mu}$ is the number of collocation points in one direction and $\nu$
  is the truncation dimension of the ANOVA decomposition, i.e. $\nu = 2$ means
  that we consider only the first and second-order interaction terms.
  For these parameters 411 points are needed for the sparse grid method and 171 points for the ANOVA method.}
  \label{tbl:sparse_anova}
	\begin{tabular}{c|c|c}
		& Sparse Grid & ANOVA \\
		\hline
		configuration & level=3 & $\mu=5, \nu=2$ \\
		\hline
		number of points & 411 & 171
	\end{tabular}
\end{table}

\begin{figure}[hp]
	\centering
	\subfigure{
		\includegraphics[width=0.45\textwidth]{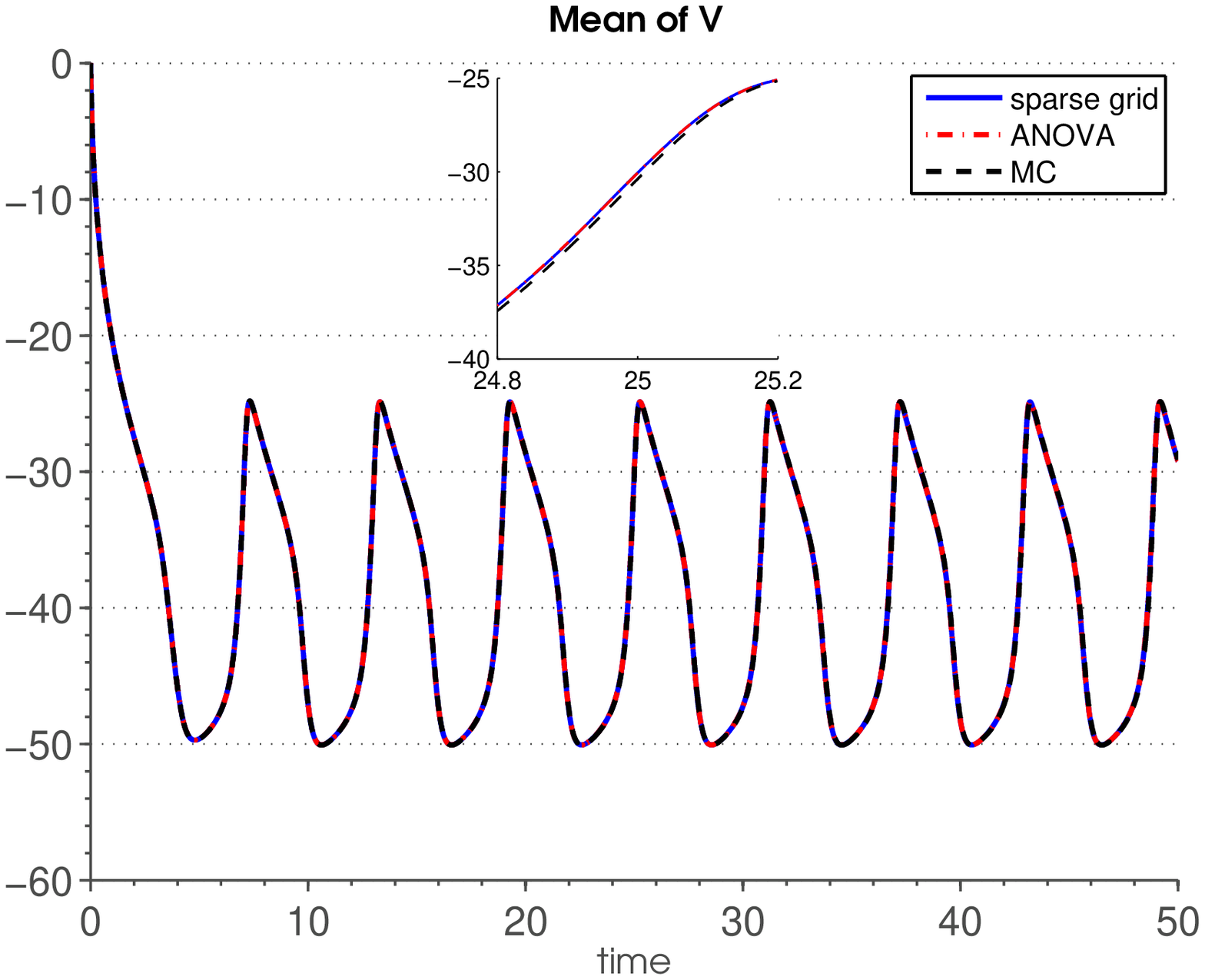} 
		\includegraphics[width=0.45\textwidth]{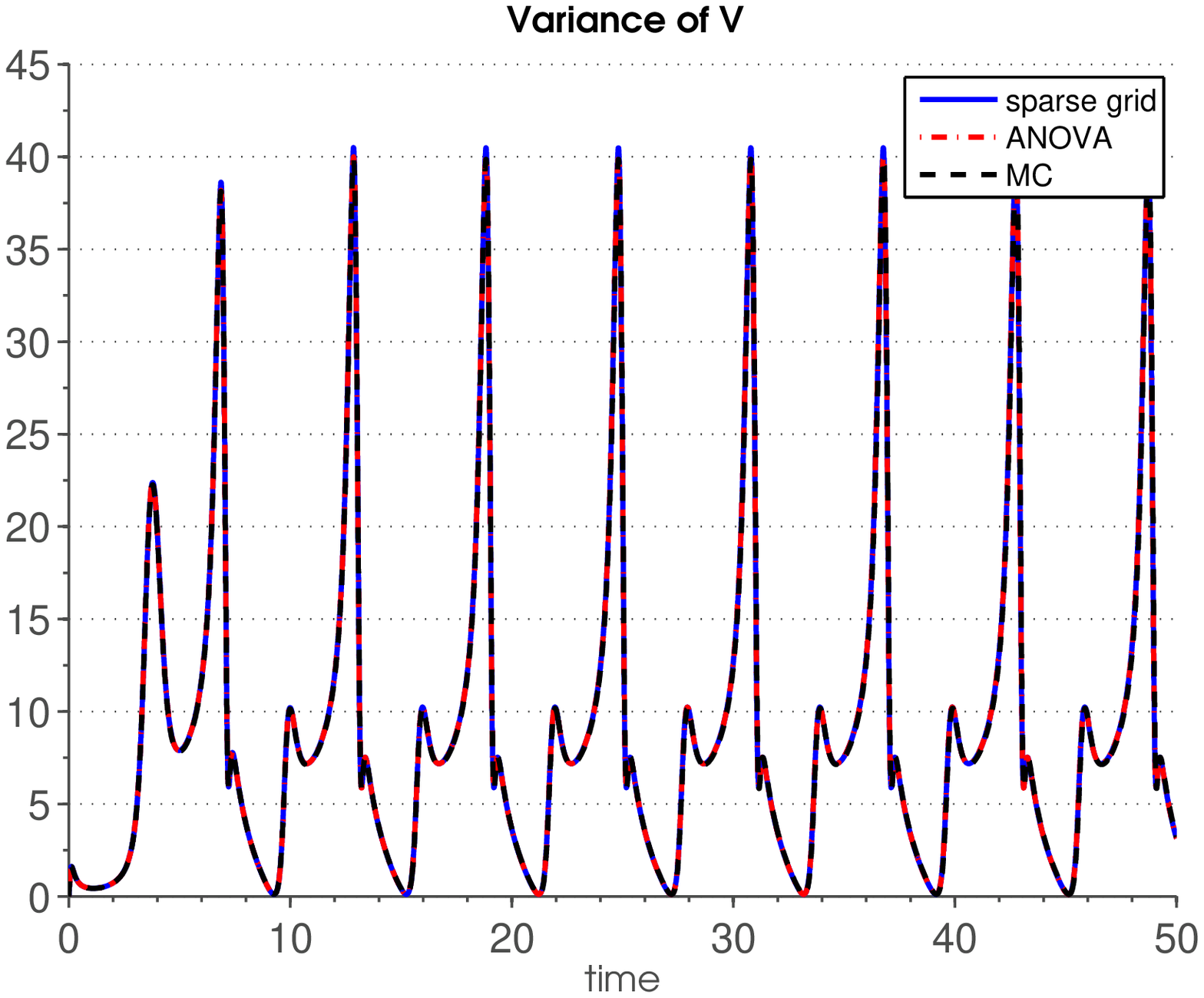}
	} 
	\caption{The mean (left) and variance (right) for $V$. MC simulations
		with 10,000 points are considered as the reference solution. Note that
	results from all methods are visually indistinguishable.}
	\label{fig:V}
\end{figure}

\begin{figure}[hp]
	\centering
	\subfigure{
		\includegraphics[width=0.45\textwidth]{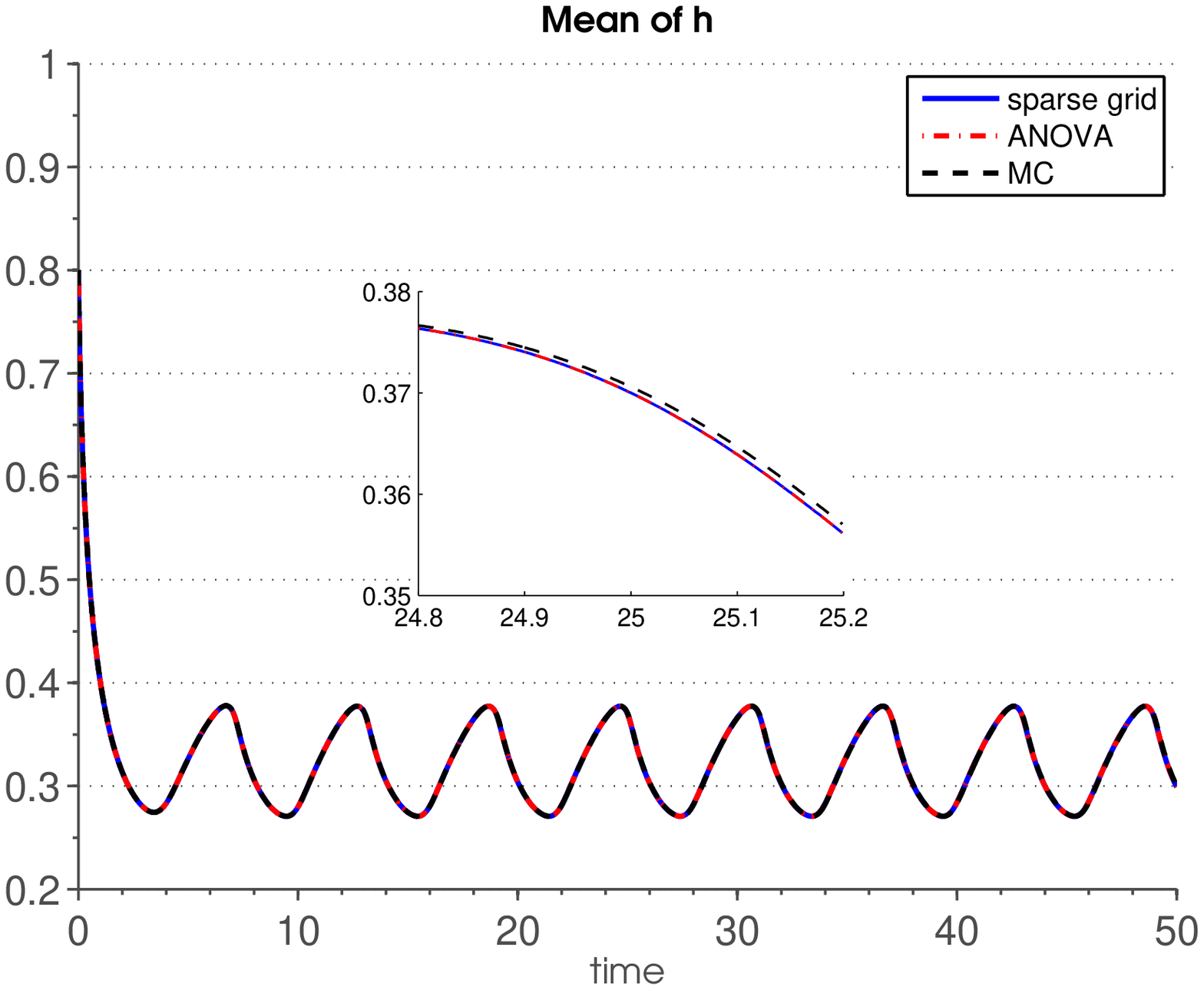} 
		\includegraphics[width=0.45\textwidth]{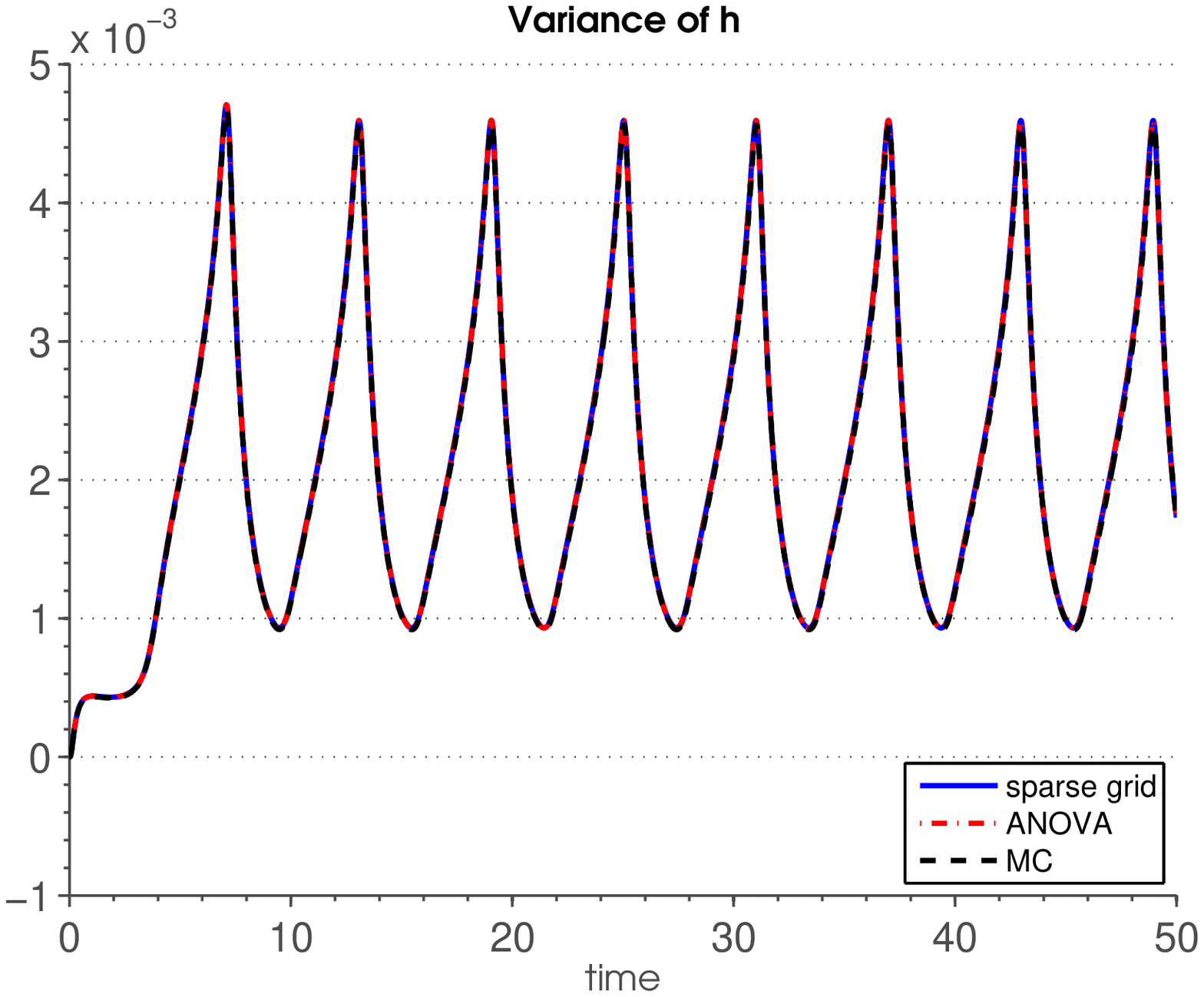} 
	} 
	\caption{The mean (left) and variance (right) for $h$. MC simulations
		with 10,000 points are considered as the reference solution. Note that
	all methods are visually indistinguishable.}
	\label{fig:h}
\end{figure}

First we solve equations (\ref{subeqn:botzinger4}) and
(\ref{subeqn:botzinger5}) for $V$ and $h$ using sparse grids,
ANOVA and MC methods and compare the mean and variance of $V$ and $h$
derived from the three methods.
For example, given
sparse grids points and corresponding weights $\{\bm \xi^{(j)},
w^{(j)}\}_{j=1}^N$, the mean and variance for $V$ can be computed as
\begin{eqnarray*}
	E[V](t) &= &\sum_{j=1}^N V(t;\bm \xi^{(j)}) w^{(j)} \\
	Var[V](t) &= &\sum_{j=1}^N V^2(t;\bm \xi^{(j)}) w^{(j)} - E[V]^2(t)
\end{eqnarray*}
where $V(t;\bm \xi^{(j)})$ is the solution to Equation
(\ref{subeqn:botzinger4}) with $\bm \xi = \bm \xi^{(j)}$.
Figures \ref{fig:V} and \ref{fig:h} show the mean and variance for $V$
and $h$, respectively, calculated using the three methods, and the
results agree well with one other. 
Note that they are visually indistinguishable
but when zoomed in (inset figure), a slight difference can be
perceived between MC and the other two methods.
This strongly suggests that the ANOVA method can help model high-dimensional
\emph{heterogeneous} parametric problems, in addition to its extensive use in
high-dimensional \emph{uncertain} parametric problems. 
Based on this observation, we consider to describe a
low-dimensional system only using the ANOVA method from now on in this
subsection.

{\bf Coarse Dynamics and Stability}. 
We will now consider the gPC coefficients $\alpha_i$ and $\beta_i, i=0,1,..M$
for $V$ and $h$, respectively as our reduced, coarse-grained
variables, i.e.~we approximately represent $V$ and $h$ as
\begin{subequations} \label{subeqn:Vh_coarse}
	\begin{align}
    V(t;\vect{\xi}) &= \sum_{i=0}^M \alpha_i(t) \Phi_i(\vect{\xi})
		\label{subeqn:V_coarse}\\
    h(t;\vect{\xi}) &= \sum_{i=0}^M \beta_i(t) \Phi_i(\vect{\xi})
		\label{subeqn:h_coarse}
	\end{align}
\end{subequations}
where each $\Phi_i(\vect{\xi}), i=0,...,M$ is a product of Legendre polynomials of
the variables in $\vect{\xi} = \{\xi_1,\xi_2,\xi_3,\xi_4\}$. 
We explore the
long-term dynamics of~\eqref{eqn:botzinger} using these coarse-grained variables and compute gPC
coefficients using ANOVA methods, as there are four heterogeneous parameters.

{\bf Equation-Free Computations}.
Availability of the governing equations for the variables of interest
is a prerequisite to modeling and computation.
However, if the underlying differential equations are
nonlinear or nontrivial and $\bm \xi$ is high-dimensional,
then the right hand side in Equation~\eqref{eq:dalpha}
is often coupled and very complicated making it almost impossible to
obtain it in explicit, closed form.
We circumvent this step
using the equation-free (EF) framework for complex, multiscale systems
modeling \cite{kevrekidis03,theodoropoulos00}. 
In this framework we
can perform system-level computational tasks without explicit
knowledge of the coarse-grained equations. 
This is accomplished through the operators that
transform between coarse and fine variables. The mapping from coarse
to fine variables is called the \emph{lifting} operator ($L$) while
the mapping from fine to coarse variables is called the
\emph{restriction} operator ($R$). 

We denote the detailed (fine), microscopic time-evolution operator
defined in Equation (\ref{eqn:botzinger_cont}) by $\phi_\tau$ (where
$\tau$
represents the number of time steps or iterations). The macroscopic
evolution operator $\Phi_\tau$ can then be defined as follows:
\begin{equation}
	\Phi_\tau(\bm \alpha(t)) = R \circ \phi_\tau \circ L(\bm \alpha(t))
\end{equation}
where $\bm \alpha(t)$ is the vector of gPC coefficients
$(\alpha_0,...,\alpha_M,\beta_0,...,\beta_M)$ in Equation
(\ref{subeqn:Vh_coarse}) representing the coarse-grained variables.
The general procedure
consists of five steps; (i) identifying observables that describe the
coarse-grained variables $\bm \alpha$, (ii) constructing a lifting operator that maps the
coarse variables to a fine scale realization, (iii) evolving the fine
scale equations for certain amount of time, (iv) restricting the resulting fine
variables to the coarse variables in order to estimate their time derivatives,
and (v) repeating the procedure to perform specific computational
tasks.

\begin{figure}[hp]
	\centering
	\subfigure{
		\includegraphics[width=0.45\textwidth]{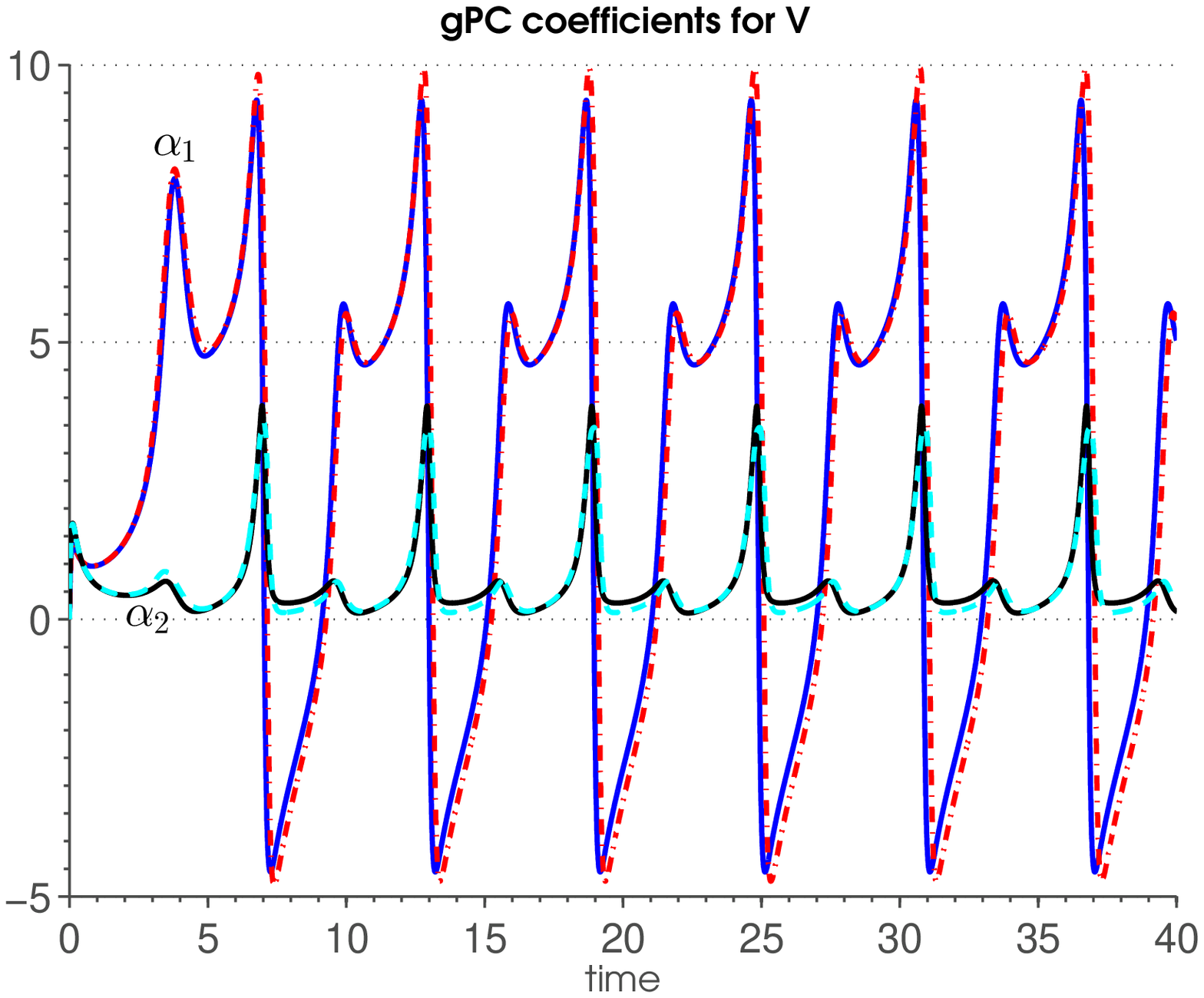} 
		\includegraphics[width=0.45\textwidth]{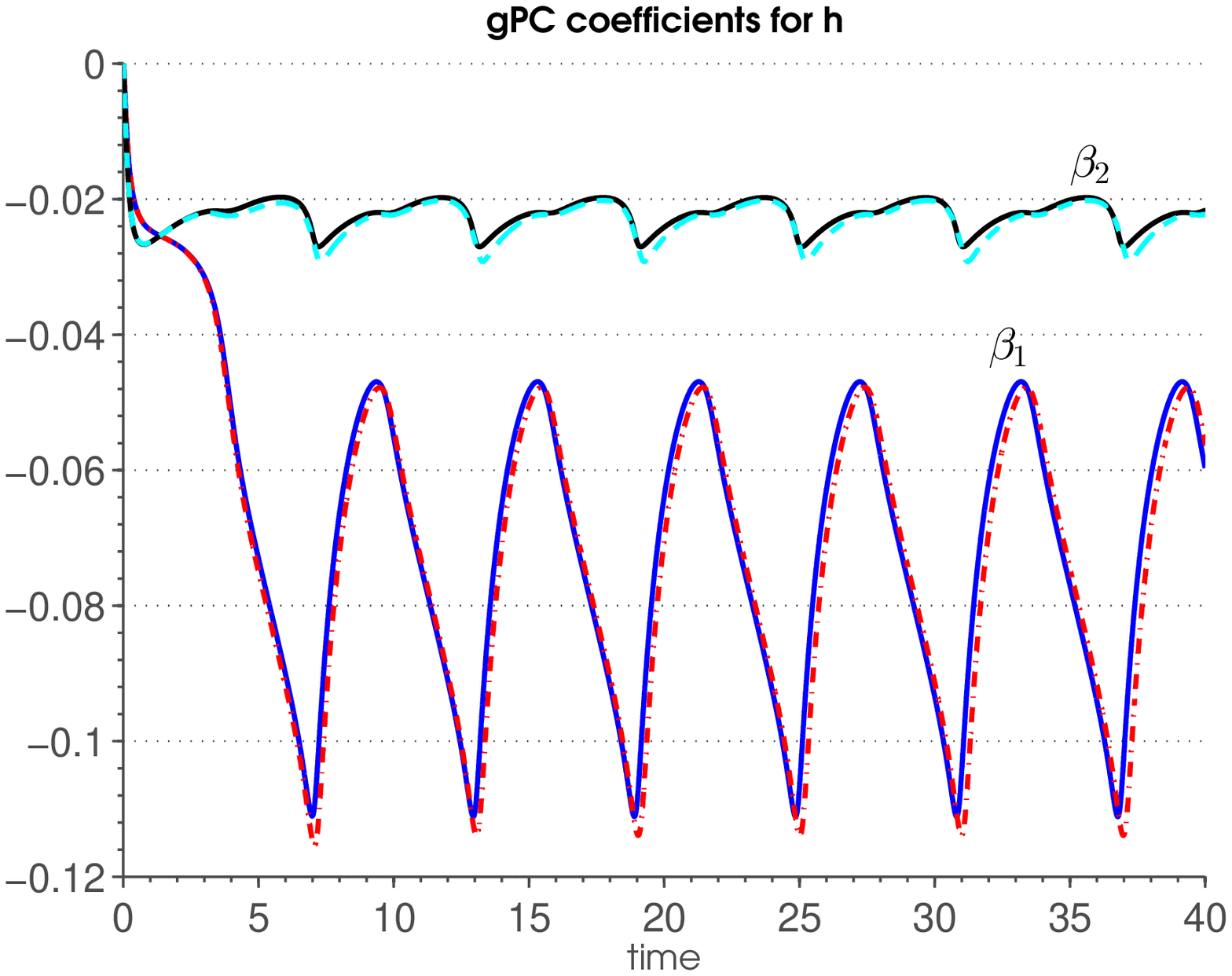} 
	} 
	\caption{Coarse projective integration (dashed lines) and detailed
		(fine) coupled dynamics (solid lines) for $V$ (left) and $h$ (right).
		Two PC coefficients $(\alpha_1,\alpha_2)$ and
		$(\beta_1,\beta_2)$ are shown for $V$ and $h$,
		respectively.
	Forward Euler with a fixed step size of $0.001$ is used as a time
	integrator. For coarse projective integration, it jumps with a
	forward Euler of 7 step after estimating time derivatives.}
	\label{fig:EF_cpi}
\end{figure}

\begin{figure}[hp]
	\centering
	\begin{tabular}{cc}
		\multirow{-3}{*}[10em]{\includegraphics[height=25em,width=.5\linewidth]{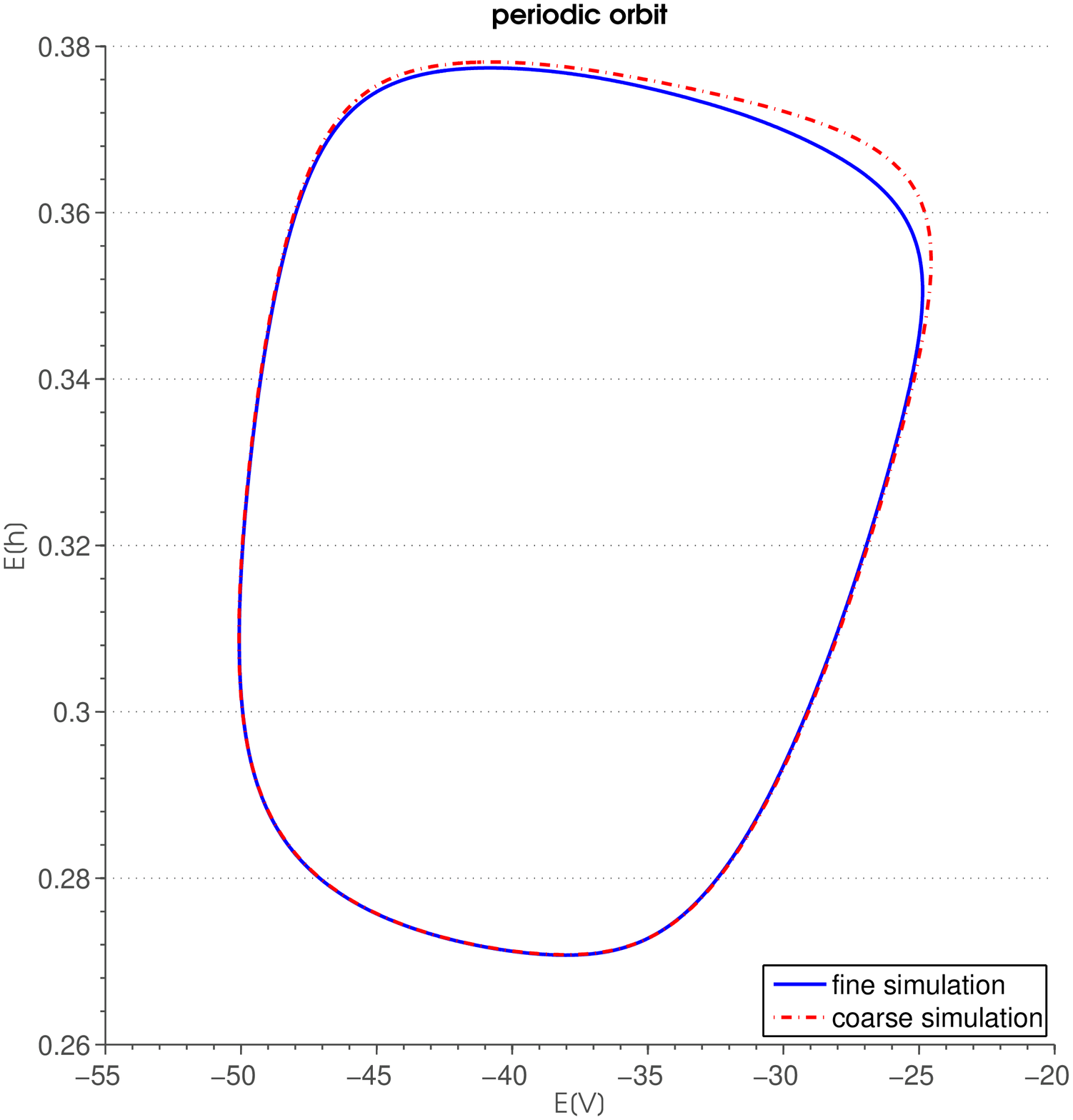}}
		&
		\includegraphics[width=.5\linewidth,height=12em]{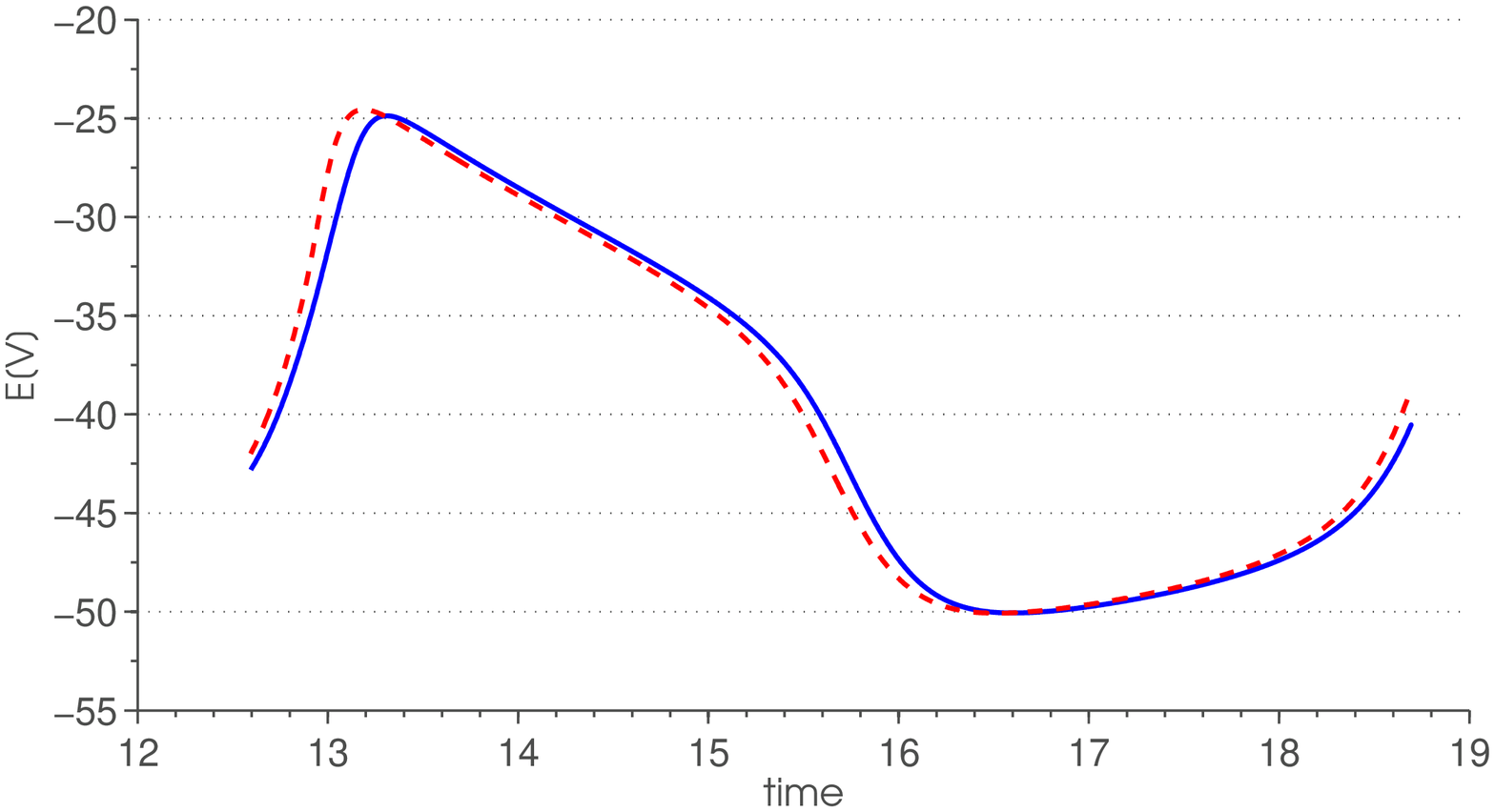}
		\\ &
		\includegraphics[width=.5\linewidth,height=12em]{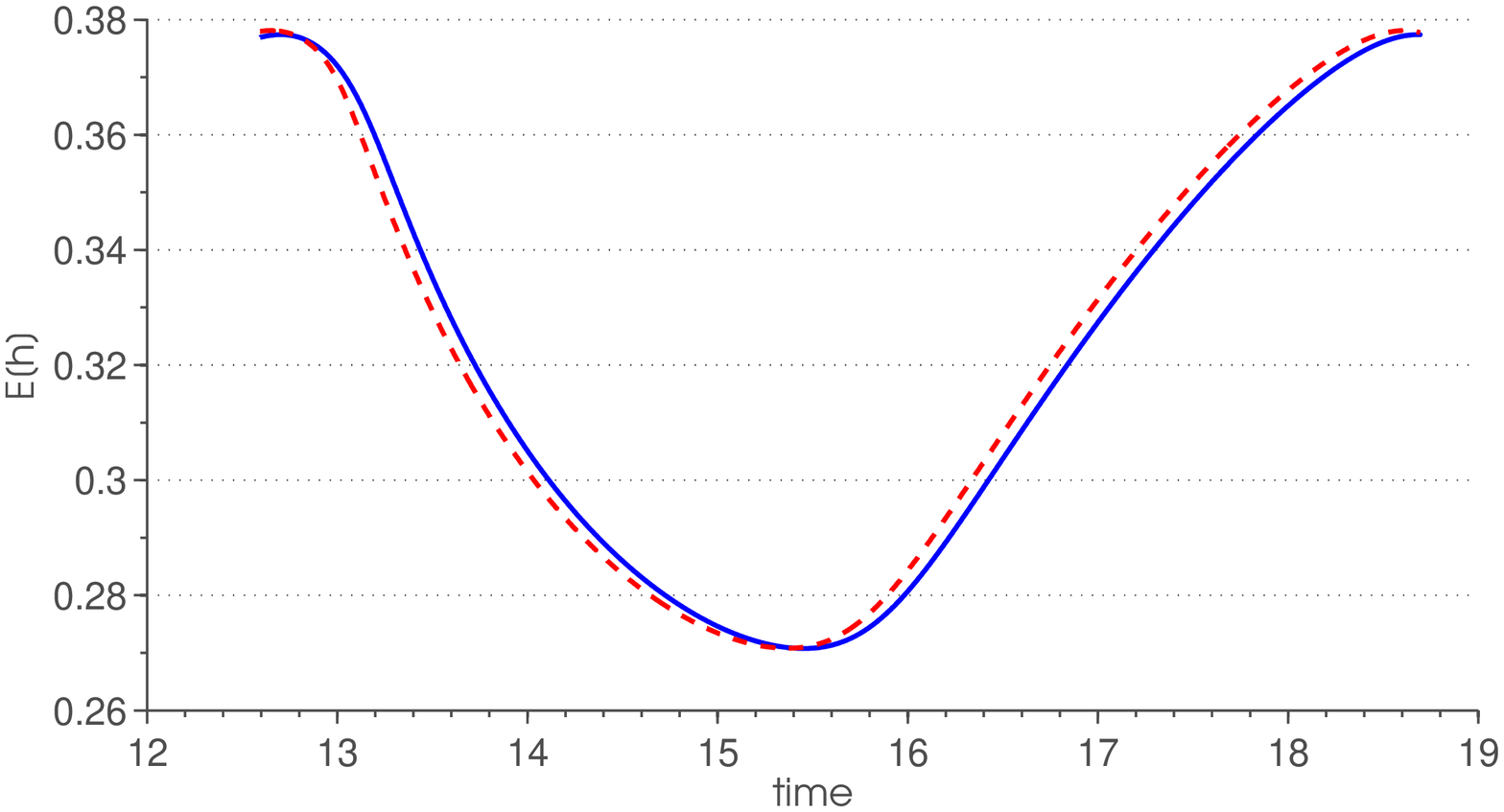}
		\\
	\end{tabular}
	\caption{Left: Periodic orbit of the mean of $V$ and the mean of $h$ when $E[I_{app}]=25$.
		Coarse projective integration (dashed) and detailed (solid)
        simulation. Temporal profile of $E[V]$ (top right) and $E[h]$
        (bottom right) corresponding to one period of limit cycle.}
	\label{fig:periodic_orbit}
\end{figure}

We first demonstrate coarse projective integration (CPI) 
\cite{gear03projective}. 
The gPC coefficients $\alpha_i$ and $\beta_i, i=0,1,..M$ for $V$ and
$h$ in Equation \eqref{subeqn:Vh_coarse} are considered as the
coarse-grained variables and obtained via Equation
(\ref{eqn:gpc_coeff}). 
For comparison, we also evolve the
detailed (fine) coupled equation (\ref{eqn:botzinger_cont}) from which
we record the coefficients (coarse-grained variables) at every time
step.
The forward Euler method with a fixed step size of $0.001$ is used as a time
integrator. For CPI, the detailed (fine) coupled system
(\ref{eqn:botzinger_cont}) is
integrated forward in time using short bursts of fine-scale
simulations consisting of 7 steps. Then, the coarse variables $\bm \alpha$ 
are evaluated according to Equation
(\ref{eqn:gpc_coeff}) where the integral is computed by the ANOVA
method given in Equation (\ref{eqn:anova_integral}). The last
few observations of the coarse variables $\bm \alpha$ are used to
estimate their time-derivative. Finally we integrate the coarse
variables with a forward Euler jump of 7 steps, thus save $7$ inner
integration steps at every $7$ steps.
Figure \ref{fig:EF_cpi} shows the second and third gPC coefficients from
coarse projective integration and from full detailed simulation, and shows that
they agree well with each other.
For the given
parameters, in particular with $E[I_{app}]=25$, the network exhibits stable, synchronized periodic
behavior as shown in Figure \ref{fig:periodic_orbit}.

\begin{figure}[h]
	\centering
	\includegraphics[width=0.8\textwidth]{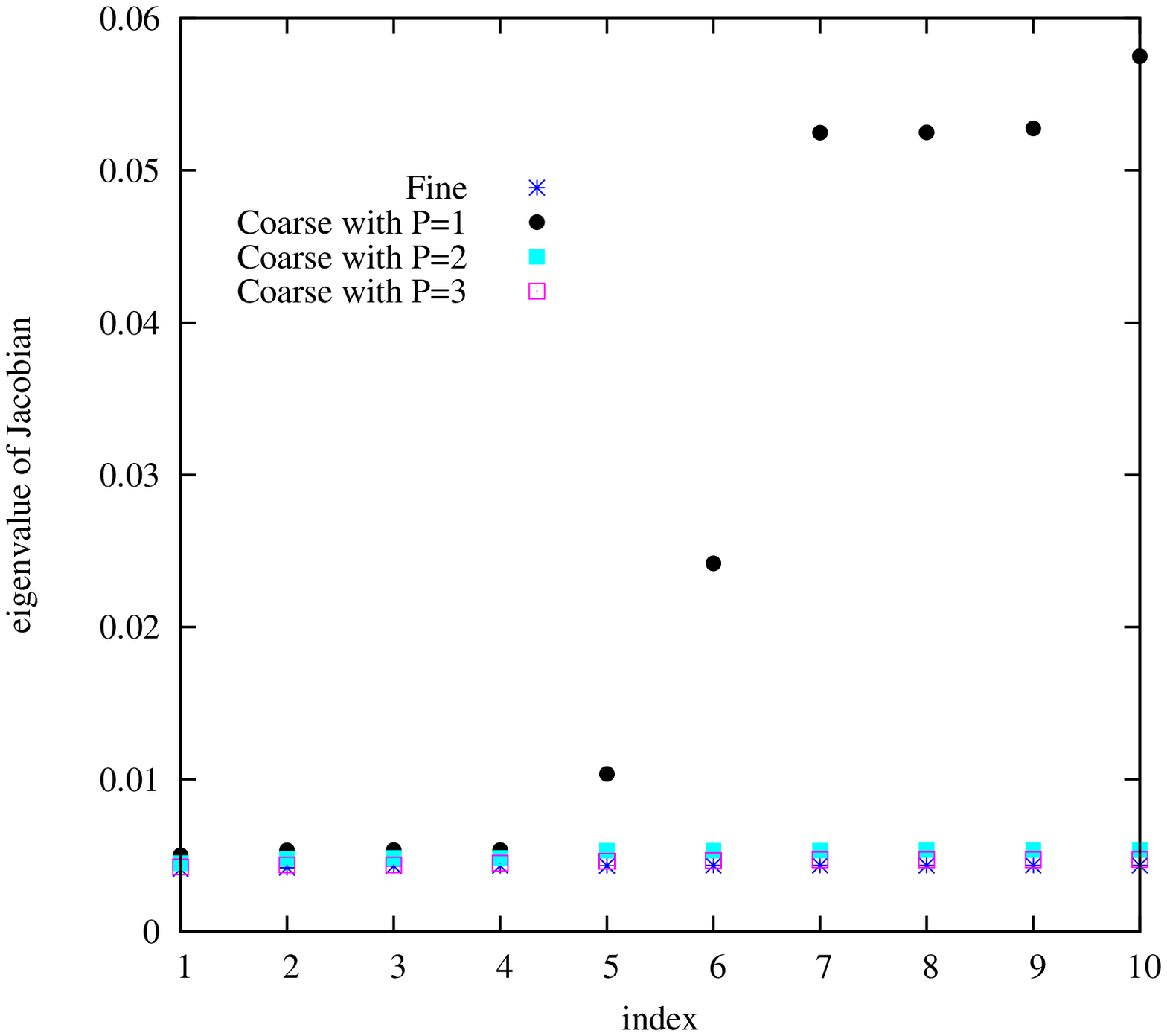} 
	\caption{Eigenvalues of the Jacobian of the fine flow map and the coarse flow map
	at corresponging fixed points, obtained with three different gPC orders: $P=1$ (leading to
	$10$ coefficients), $P=2$ ($30$
	coefficients), and $P=3$ (70 coefficients). As the polynomial
	degree -and thus the number of coarse variables- increases, the eigenvalues
	from the coarse flow simulation show increasingly better agreement with those from
	the fine simulation.}
	\label{fig:eigenvalues_fine_coarse}
\end{figure}

The equation-free approach is also useful for computing long-time
(stationary) states and their stability and dependence on parameters
\cite{kevrekidis03,moon05}. 
The coarse time-stepper $\Phi_{\tau}(\bm
\alpha(t))$ is defined as mapping from $\bm \alpha(t)$ to $\bm
\alpha(t+\tau)$ via one iteration of the equation-free method as
mentioned in the above: lifting a coarse-grained initial condition
$\bm \alpha(t)$ to one or more consistent fine initial conditions, 
integrating the full (fine) model for a (short) time $\tau$, and then restricting to
the coarse observable of the final fine state $\Phi_{\tau}$. 
In order to compute the stationary states we solve for the fixed point $\bm \alpha^*$
satisfying
\begin{equation}
	 F_{\tau}(\bm \alpha) \equiv \Phi_{\tau}(\bm \alpha) - \bm \alpha = 0,
\end{equation}
which is referred to as the coarse flow map.
Iterative matrix-free linear algebra algorithms such as Newton-GMRES can be used
to find zeros of such a function in the absence of explicit equations
for the dynamics of the coarse variables $\bm \alpha$.
Eigenvalues of the Jacobian of the coarse flow map $F_{\tau}$ evaluated at a fixed point reveal
the (coarse grained) stability of that fixed point and help determine the nature of
its potential bifurcations.
Figure \ref{fig:eigenvalues_fine_coarse}
shows the first 10 eigenvalues of the Jacobians of both the fine and coarse flow maps at
equivalent fixed points. 
As the polynomial degree in the coarse flow map (the
number of coarse variables)
increases, these coarse eigenvalue estimates are expected to approach
the leading eigenvalues of the Jacobian of the fine flow map, and this is clearly
seen in Figure \ref{fig:eigenvalues_fine_coarse}.

\subsection{Case II: intrinsic \emph{and} structural heterogeneity}
\label{sec:struc}
We consider Equation (\ref{eqn:botzinger}) with the following physiological parameter values~\cite{carlo12}
\begin{equation}
  V_{N_a} = 50, \quad V_{syn} = 0, \quad g_{syn} = 0.3,
  \quad g_l = 2.4, \quad V_l = - 65, \quad \varepsilon = 0.1,
  \quad C = 0.21. \nonumber
\end{equation}
For a heterogeneous network, $I_{app}$ is chosen to follow a
uniform distribution on $[17.5,32.5]$, parameterized by $I_{app} =
25+7.5\omega$ where $\omega$ is uniformly distributed on $[-1,1]$.
Neurons are connected in a Chung-Lu type network \cite{carlo12chimeras}, i.e. neurons
$i$ and $j$ are
connected (i.e.~$A_{ij}=1$) with probability
\begin{equation}
	p_{ij} = \min{\left(\frac{\phi_i \phi_j}{\sum_k \phi_k},1\right)}
\end{equation}
 where $\phi_i = p N (i/N)^r, i=1,...,N$ and $N$ is
the number of neurons. We choose $N=512, p=0.5$ and $r=0.1$.

Clearly, we can consider the way the neurons are connected in the network as
a different type of heterogeneity: a \emph{structural} heterogeneity, where 
neurons are connected between them in different ways, as opposed to neurons
having different individual parameters (an \emph{intrinsic} heterogeneity, of the
type we have discussing up to now).
In our case, we assume that this structural heterogeneity is well described by the
degree distribution: the degree of each neuron denoted by $\kappa$ is the important structural 
heterogeneous parameter, and its probability distribution is the 
degree distribution of our Chung-Lu network. 

For these parameter values, and a particular realization of a Chung-Lu
network with 512
neurons, we observe that the network eventually synchronizes, and all
neurons evolve along
a periodic trajectory (each in a slightly different periodic path,
since the neurons differ both
intrinsically and in their connectivities). 
At any point in time, the
state at each neuron, $(V_i,\,h_i)(t)$ can be approximated by a smooth
surface in two heterogeneous parameters
$\vect{\xi}=(\kappa_i,I_{\mathrm{app},i})_{i=1}^N$ according to Equation
(\ref{subeqn:Vh_coarse}). These are parameters in the sense that they do not change in
time--they are still unique for each neuron.
If indeed the behavior can be expressed as a function of our two heterogeneous parameters
and time, this suggests that at every moment in time the values of the dynamic
variables of each neuron would lie on a smooth surface, here a two dimensional one,
parametrized by the two measures of heterogeneity.
At every point in time the $512$ individual variable values, one for every neuron,
would lie on, or very close to, this surface.

\begin{figure}[h]
	\begin{tabular}{cc}
		\multirow{-3}{*}[10em]{\includegraphics[height=25em,width=.5\linewidth]{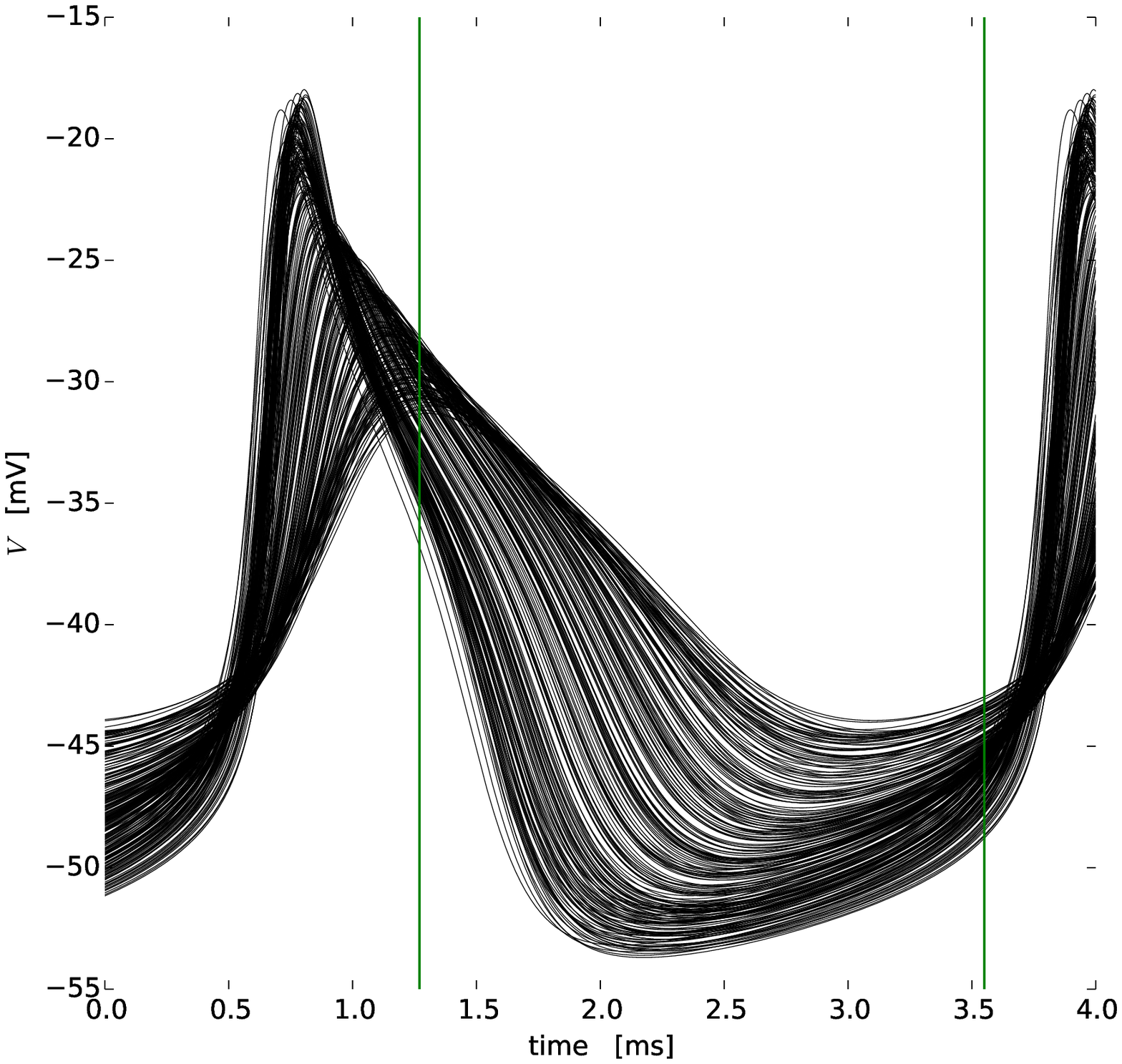}}
		&
		\includegraphics[width=.5\linewidth,height=10em]{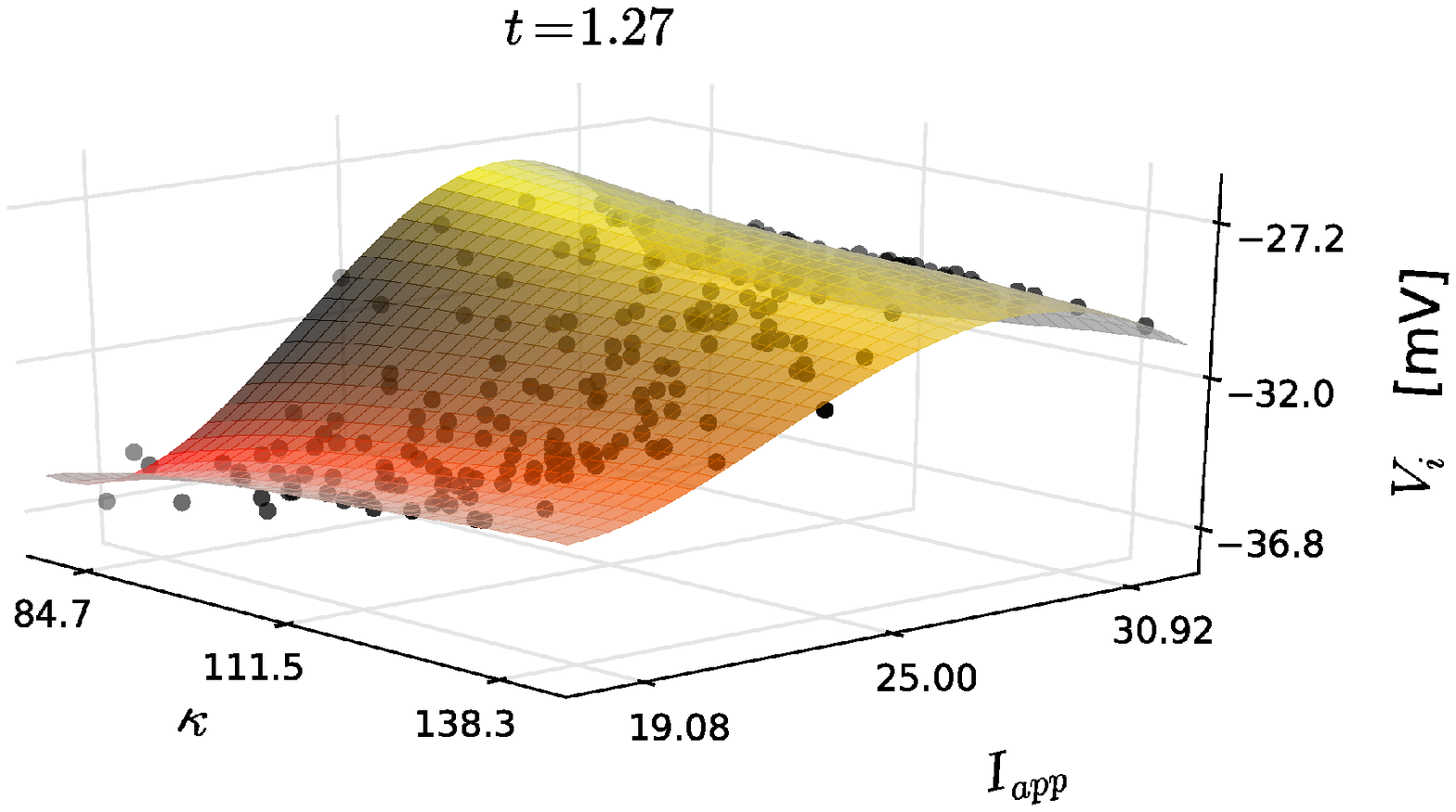}
		\\ &
		\includegraphics[width=.5\linewidth,height=10em]{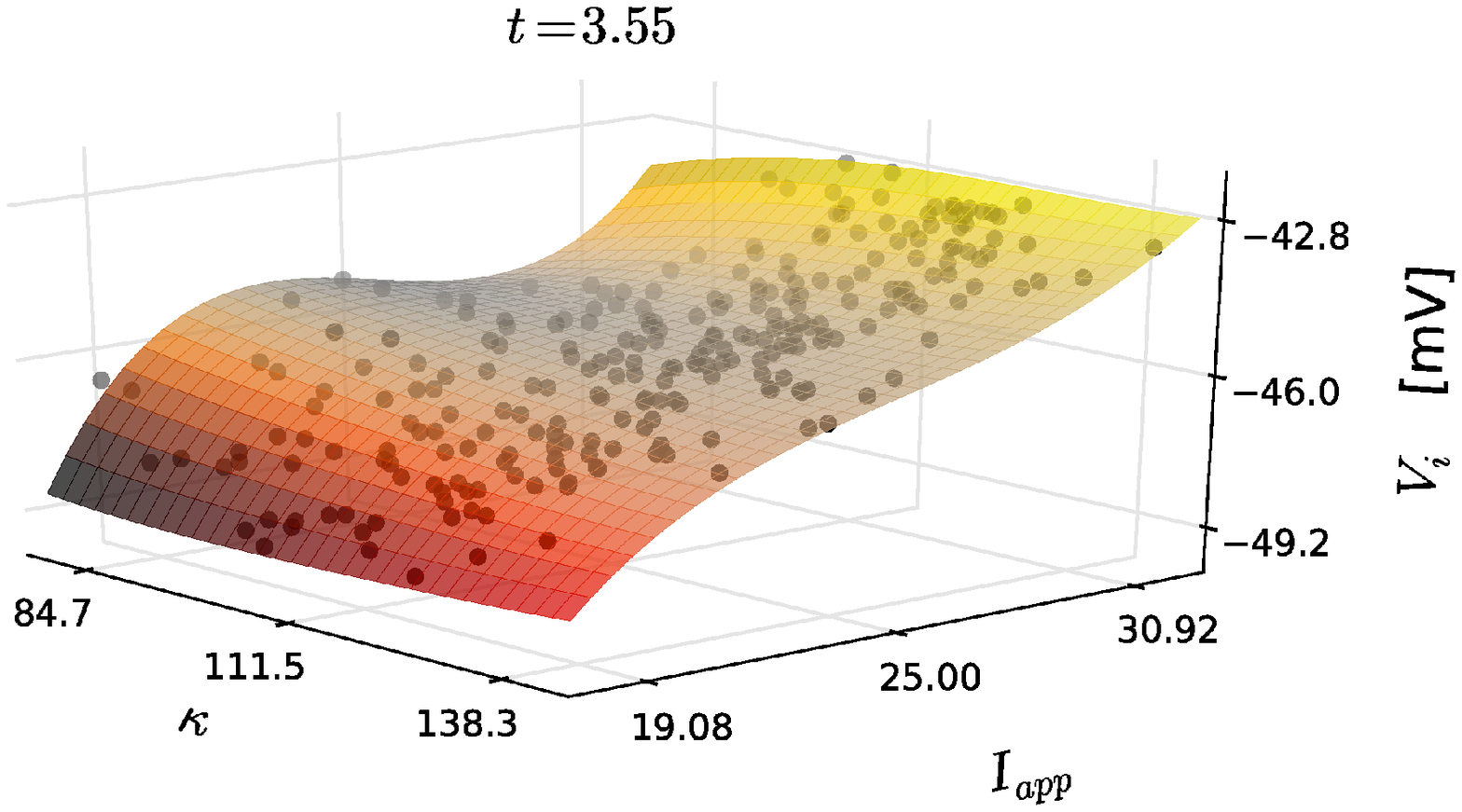}
		\\
	\end{tabular}
  \caption{
        \textit{Left:} Potential $(V_i)$ in Equation (\ref{subeqn:botzinger1}) with two
		heterogeneous parameters $(\kappa_i, I_{\mathrm{app},i})$.
        Surface of the $V_i$ as a function of the two parameters
		$(\kappa_i, I_{\mathrm{app},i})$ at $t=1.27$ (\textit{top right}) and
        $t=3.55$ (\textit{bottom right}).
        The neurons lie on, or very close to, the smooth manifold.
        Initial conditions for the integration of \eqref{subeqn:botzinger1}
        at $t=0$ are
		a point $\vec X(t=0;\,\vect{\xi}) = \{ \vec V, \vec h \} (t=0;\, \{ \kappa,I_{app} \} )$ found by simple forward integration
        to be on or very close to the attracting limit cycle.
        In the left panel, we show approximately one period of this limit cycle,
        observed in the 512 $V$ traces over time.
        }
	\label{fig:hettraj_surf}
\end{figure}

Figure
\ref{fig:hettraj_surf} shows the potential $V_i$ of all the neurons 
for $0\leq t\leq 4$ and the evolving ``heterogeneity surface" of the potential $V_i$ 
at two instances in time $t=1.27$ and $3.55$ (marked on the figure) as a function of the two
heterogeneous parameters, which are randomly picked at on a limit
cycle.
The fact that, for all practical purposes, the values of the variables for each neuron lie on, or
close to such a
smooth surface,  implies that a gPC representation performs well  
as a coarse-grained descriptor of the heterogeneous neuronal population.

\begin{figure}[h]
	\centering
	\includegraphics[width=\textwidth]{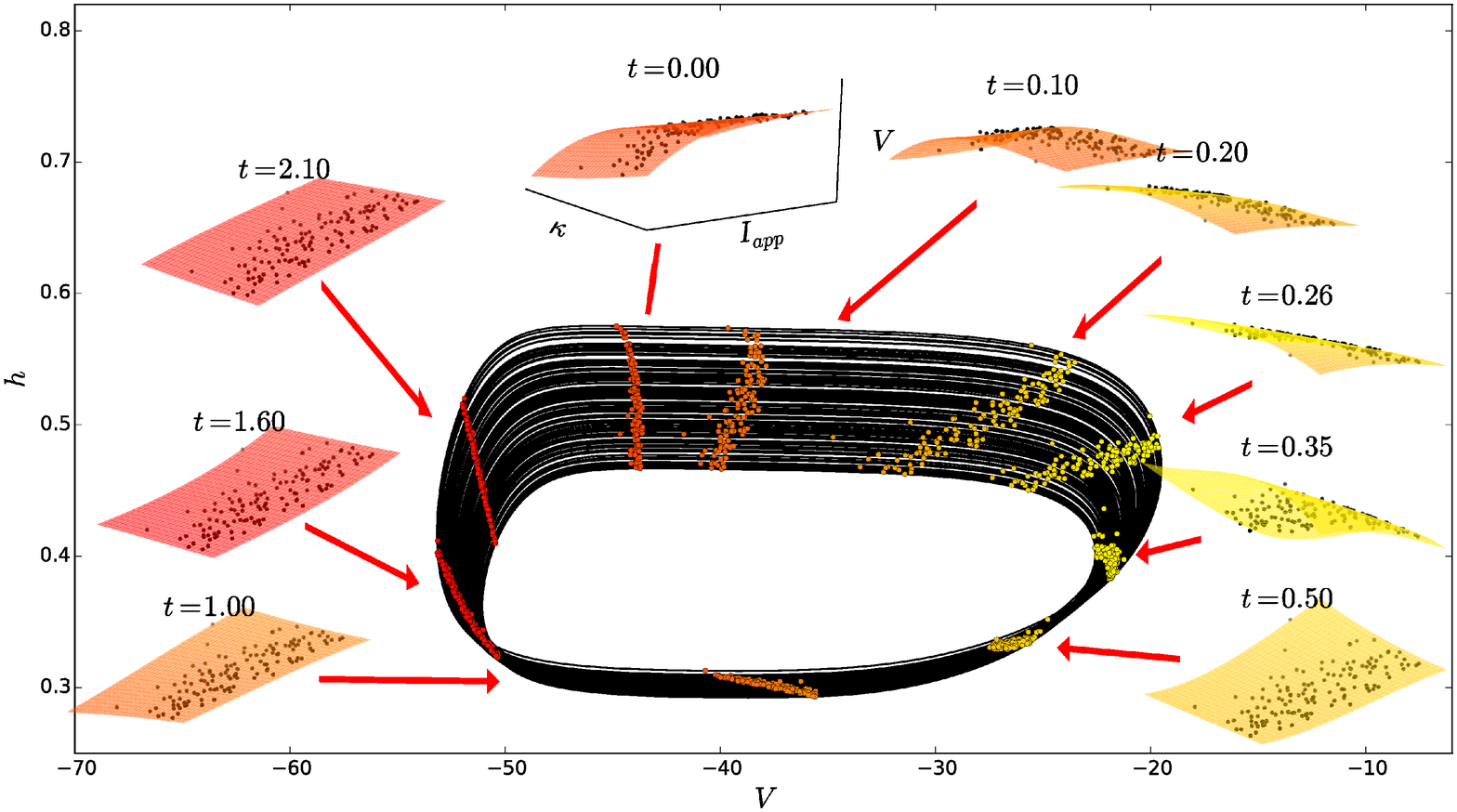} 
    \caption{Limit cycles of the $(V_i,h_i)$ for $N=512$ neurons. Each filled
        circle represents the potential of one neuron, with different colors denoting different time
        snapshots along the synchronized oscillation.
        The oscillations proceed in the clockwise direction, and the surfaces
        in the insets show the $V_i$ as functions of the two heterogeneous
        parameters at nine different times cut.}
	\label{fig:limitcycle}
\end{figure}

Figure \ref{fig:limitcycle} shows a phase portrait view of the limit cycle synchronized 
oscillation for all the neurons. In the
insets we show, at seven different time instances, the potential V of each neurons (represented by colored filled circles),
clearly lying on, or very close to, the smooth two-dimensional surface of the coarse-grained description.

%
%
\section{Conclusion}
\label{sec:conclusion}
We have proposed and demonstrated the use of several distinct forms of dimension reduction
for the computationally efficient study of heterogeneous networks of
coupled neurons.
In Case I we considered an all-to-all coupled network with four
independent heterogeneous parameters. 
To efficiently simulate such a
network we need to approximate a four-dimensional integral, which we
accomplished using ANOVA methods. 
A reduced model of this type of network can
also be formulated using  coefficients in a polynomial chaos expansion
in the heterogeneous parameters as the ``coarse'' variables.
Having such a reduced model leads to an improvement in the speed for a variety of
computations of interest (direct simulation, coarse limit cycle computation,
coarse stability analysis) which we demonstrated using the equation-free framework.

In Case II we considered a network with \emph{both}
intrinsic and structural heterogeneity, and showed that we could expand the state variables
in polynomials of both the intrinsically varying parameter and a feature of the network connectivity -
in this case, the degree of each neuron.
To do this, we need to construct orthogonal polynomials with respect to the network degree distribution.
If this (integer) distribution is known \emph{a priori}, then the polynomials can be found in the
literature~\cite{xiu10},
or easily constructed using the recurrence relation~\cite{wan06}.
If the distribution is unknown, and we only have samples of it available, then the convergence of the
``empirical" polynomials based on the sampled distributions, to the ``true" distributions
at the limit of infinite neurons becomes an interesting research
problem that we are currently investigating.
We believe that all these approaches can play an important practical role in accelerating the
computational study (and, in general, the modeling) of complex heterogeneous networks,
and we are exploring the practical limits of (a) how many independently distributed 
heterogeneous parameters one can usefully approximate and (b) the modeling of 
heterogeneities that are not independently distributed, but rather exhibit correlations.

%
%

{\bf Acknowledgements}. This work was partially supported by the US National Science Foundation
and by the US AFOSR. The hospitality and support of the Institute for Advanced Study at the T. U. 
Muenchen, where I.G.K was a Hans Fischer Senior Fellow, and C.L. a visitor is gratefully acknowledged.

%
%
\bibliography{equations}

\begin{thebibliography}{10}

\bibitem{ashwin92}
Peter Ashwin and James~W Swift.
\newblock The dynamics of $n$ weakly coupled identical oscillators.
\newblock {\em Journal of Nonlinear Science}, 2(1):69--108, 1992.

\bibitem{bold07}
Katherine~A Bold, Yu~Zou, Ioannis~G Kevrekidis, and Michael~A Henson.
\newblock An equation-free approach to analyzing heterogeneous cell population
  dynamics.
\newblock {\em Journal of mathematical biology}, 55(3):331--352, 2007.

\bibitem{butera99models_a}
Robert~J Butera, John Rinzel, and Jeffrey~C Smith.
\newblock Models of respiratory rhythm generation in the pre-b{\"o}tzinger
  complex. i. bursting pacemaker neurons.
\newblock {\em Journal of neurophysiology}, 82(1):382--397, 1999.

\bibitem{butera99models_b}
Robert~J Butera, John Rinzel, and Jeffrey~C Smith.
\newblock Models of respiratory rhythm generation in the pre-b{\"o}tzinger
  complex. ii. populations of coupled pacemaker neurons.
\newblock {\em Journal of Neurophysiology}, 82(1):398--415, 1999.

\bibitem{rubin10optimal}
Justin~R Dunmyre and Jonathan~E Rubin.
\newblock Optimal intrinsic dynamics for bursting in a three-cell network.
\newblock {\em SIAM Journal on Applied Dynamical Systems}, 9(1):154--187, 2010.

\bibitem{fisher}
R.~Fisher.
\newblock {\em Statistical Methods for Research Workers}.
\newblock Oliver and Boyd, 1925.

\bibitem{foo08mepcm}
Jasmine Foo, Xiaoliang Wan, and George~Em Karniadakis.
\newblock The multi-element probabilistic collocation method (me-pcm): Error
  analysis and applications.
\newblock {\em Journal of Computational Physics}, 227(22):9572--9595, 2008.

\bibitem{foo10mepcm}
Jasmine~Y. Foo and George~Em Karniadakis.
\newblock Multi-element probabilistic collocation in high dimensions.
\newblock {\em Journal of Computational Physics}, 229:1536--1557, 2010.

\bibitem{gear03projective}
C~William Gear and Ioannis~G Kevrekidis.
\newblock Projective methods for stiff differential equations: problems with
  gaps in their eigenvalue spectrum.
\newblock {\em SIAM Journal on Scientific Computing}, 24(4):1091--1106, 2003.

\bibitem{griebel98sparse}
Thomas Gerstner and Michael Griebel.
\newblock Numerical integration using sparse grids.
\newblock {\em Numerical algorithms}, 18(3-4):209--232, 1998.

\bibitem{ghanem03}
Roger~G Ghanem and Pol~D Spanos.
\newblock {\em Stochastic finite elements: a spectral approach}.
\newblock Courier Corporation, 2003.

\bibitem{hassard78}
Brian Hassard.
\newblock Bifurcation of periodic solutions of the hodgkin-huxley model for the
  squid giant axon.
\newblock {\em Journal of Theoretical Biology}, 71(3):401--420, 1978.

\bibitem{hoeffding}
W.~Hoeffding.
\newblock A class of statistics with asymptotically normal distributions.
\newblock {\em Annals of Math. Statist.}, 19:293--325, 1948.

\bibitem{kevrekidis03}
Ioannis~G Kevrekidis, C~William Gear, James~M Hyman, Panagiotis~G Kevrekidis,
  Olof Runborg, Constantinos Theodoropoulos, et~al.
\newblock Equation-free, coarse-grained multiscale computation: Enabling
  mocroscopic simulators to perform system-level analysis.
\newblock {\em Communications in Mathematical Sciences}, 1(4):715--762, 2003.

\bibitem{laing08}
Carlo~R Laing and Ioannis~G Kevrekidis.
\newblock Periodically-forced finite networks of heterogeneous globally-coupled
  oscillators: a low-dimensional approach.
\newblock {\em Physica D: Nonlinear Phenomena}, 237(2):207--215, 2008.

\bibitem{carlo12chimeras}
Carlo~R Laing, Karthikeyan Rajendran, and Ioannis~G Kevrekidis.
\newblock Chimeras in random non-complete networks of phase oscillators.
\newblock {\em Chaos: An Interdisciplinary Journal of Nonlinear Science},
  22(1):013132, 2012.

\bibitem{carlo12}
Carlo~R Laing, Yu~Zou, Ben Smith, and Ioannis~G Kevrekidis.
\newblock Managing heterogeneity in the study of neural oscillator dynamics.
\newblock {\em Journal of mathematical neuroscience}, 2(1):5, 2012.

\bibitem{moon15}
Sung~Joon Moon, Katherine~A Cook, Karthikeyan Rajendran, Ioannis~G Kevrekidis,
  Jaime Cisternas, and Carlo~R Laing.
\newblock Coarse-grained clustering dynamics of heterogeneously coupled
  neurons.
\newblock {\em The Journal of Mathematical Neuroscience (JMN)}, 5(1):1--20,
  2015.

\bibitem{moon05}
Sung~Joon Moon, R.~Ghanem, and I.~G. Kevrekidis.
\newblock Coarse graining the dynamics of coupled oscillators.
\newblock {\em Phys. Rev. Lett.}, 96:144101, Apr 2006.

\bibitem{rubin02}
Jonathan Rubin and David Terman.
\newblock Synchronized activity and loss of synchrony among heterogeneous
  conditional oscillators.
\newblock {\em SIAM Journal on Applied Dynamical Systems}, 1(1):146--174, 2002.

\bibitem{rubin06bursting}
Jonathan~E Rubin.
\newblock Bursting induced by excitatory synaptic coupling in nonidentical
  conditional relaxation oscillators or square-wave bursters.
\newblock {\em Physical Review E}, 74(2):021917, 2006.

\bibitem{sobol01}
I.M. Sobol'.
\newblock Global sensitivity indices for nonlinear mathematical models and
  their monte carlo estimates.
\newblock {\em Mathematics and Computers in Simulation}, 55:271--280, 2001.

\bibitem{theodoropoulos00}
Constantinos Theodoropoulos, Yue-Hong Qian, and Ioannis~G Kevrekidis.
\newblock “coarse” stability and bifurcation analysis using time-steppers:
  A reaction-diffusion example.
\newblock {\em Proceedings of the National Academy of Sciences},
  97(18):9840--9843, 2000.

\bibitem{wan06}
Xiaoliang Wan and George~Em Karniadakis.
\newblock Beyond wiener--askey expansions: handling arbitrary pdfs.
\newblock {\em Journal of Scientific Computing}, 27(1-3):455--464, 2006.

\bibitem{xiu10}
Dongbin Xiu.
\newblock {\em Numerical methods for stochastic computations: a spectral method
  approach}.
\newblock Princeton University Press, 2010.

\bibitem{xiu05pcm}
Dongbin Xiu and Jan~S Hesthaven.
\newblock High-order collocation methods for differential equations with random
  inputs.
\newblock {\em SIAM Journal on Scientific Computing}, 27(3):1118--1139, 2005.

\bibitem{xiu02gpc}
Dongbin Xiu and George~Em Karniadakis.
\newblock The wiener--askey polynomial chaos for stochastic differential
  equations.
\newblock {\em SIAM journal on scientific computing}, 24(2):619--644, 2002.

\bibitem{choi12anova}
Xiu Yang, Minseok Choi, Guang Lin, and George~Em Karniadakis.
\newblock Adaptive anova decomposition of stochastic incompressible and
  compressible flows.
\newblock {\em Journal of Computational Physics}, 231(4):1587--1614, 2012.

\end{thebibliography}
\bibliographystyle{plain}

\end{document}